%% file: main.tex
\documentclass[twocolumn]{aastex631}

\usepackage{color}
\usepackage{graphicx, epstopdf} 
\usepackage{gensymb}
\usepackage{natbib}
\usepackage{amsmath}
\usepackage{amsfonts}
\usepackage{amsthm}
\usepackage{xspace}
\usepackage{multirow}
\usepackage{url}

\received{}
\revised{}
\accepted{}
\submitjournal{ApJ}

\shorttitle{GBT Timing of Transient Pulsars}
\shortauthors{Lewis et al.}

\graphicspath{{./}{figures/}}

\begin{document}

\title{Multi-Method Timing of Transient Radio Pulsars with the GBT350 and GBNCC Surveys}

\correspondingauthor{E.~F.~Lewis}
\email{efl0003@mix.wvu.edu}

\author[0000-0002-2972-522X]{E.~F.~Lewis}
\affiliation{West Virginia University, Department of Physics and Astronomy, P. O. Box 6315, Morgantown, WV 26506, USA}
\affiliation{Center for Gravitational Waves and Cosmology, West Virginia University, Chestnut Ridge Research Building, Morgantown, WV 26506, USA}

\author[0000-0001-7697-7422]{M.~A.~McLaughlin}
\affiliation{West Virginia University, Department of Physics and Astronomy, P. O. Box 6315, Morgantown, WV 26506, USA}
\affiliation{Center for Gravitational Waves and Cosmology, West Virginia University, Chestnut Ridge Research Building, Morgantown, WV 26506, USA}

\author[0000-0002-1075-3837]{J.~K.~Swiggum}
\affiliation{Department of Physics, Lafayette College, Easton, PA 18042, USA}

\author[0000-0003-4046-884X]{H.~Blumer}
\affiliation{West Virginia University, Department of Physics and Astronomy, P. O. Box 6315, Morgantown, WV 26506, USA}
\affiliation{Center for Gravitational Waves and Cosmology, West Virginia University, Chestnut Ridge Research Building, Morgantown, WV 26506, USA}

\author{J.~Boyles}
\affiliation{Department of Physics and Astronomy, Western Kentucky University, 1906 College Heights Blvd., Bowling Green, KY 42101, USA}

\author[0000-0002-3426-7606]{P.~Chawla}
\affiliation{Anton Pannekoek Institute for Astronomy, University of Amsterdam, Science Park 904, 1098 XH Amsterdam, The Netherlands}
\affiliation{ASTRON, Netherlands Institute for Radio Astronomy, Oude Hoogeveensedijk 4, 7991 PD Dwingeloo, The Netherlands}

\author[0000-0001-8885-6388]{T.~Dolch}
\affiliation{Department of Physics, Hillsdale College, 33 E. College Street, Hillsdale, Michigan 49242, USA}
\affiliation{Eureka Scientific, 2452 Delmer Street, Suite 100, Oakland, CA 94602-3017, USA}

\author[0000-0003-2317-1446]{J.~W.~T.~Hessels}
\affiliation{Anton Pannekoek Institute for Astronomy, University of Amsterdam, Science Park 904, 1098 XH Amsterdam, The Netherlands}
\affiliation{ASTRON, Netherlands Institute for Radio Astronomy, Oude Hoogeveensedijk 4, 7991 PD Dwingeloo, The Netherlands}
\affiliation{Trottier Space Institute, McGill University, 3550 rue University, Montréal, QC H3A 2A7, Canada}
\affiliation{Department of Physics, McGill University, 3600 rue University, Montréal, QC H3A 2T8, Canada}

\author[0000-0001-6295-2881]{D.~L.~Kaplan}
\affiliation{Center for Gravitation, Cosmology and Astrophysics, Department of Physics and Astronomy, University of Wisconsin-Milwaukee, P.O. Box 413, Milwaukee, WI 53201, USA}

\author{C.~Karako-Argaman}
\affiliation{Department of Physics, McGill University, 3600 rue University, Montréal, QC H3A 2T8, Canada}

\author[0000-0001-9345-0307]{V.~Kaspi}
\affiliation{Trottier Space Institute, McGill University, 3550 rue University, Montréal, QC H3A 2A7, Canada}
\affiliation{Department of Physics, McGill University, 3600 rue University, Montréal, QC H3A 2T8, Canada}

\author[0000-0001-8864-7471]{V.~Kondratiev}
\affiliation{ASTRON, Netherlands Institute for Radio Astronomy, Oude Hoogeveensedijk 4, 7991 PD Dwingeloo, The Netherlands}

\author[0000-0002-2034-2986]{L.~Levin}
\affiliation{Jodrell Bank Centre for Astrophysics, School of Physics and Astronomy, University of Manchester, Manchester M13 9PL, UK}

\author[0000-0001-5229-7430]{R.~S.~Lynch}
\affiliation{Green Bank Observatory, P.O. Box 2, Green Bank, WV 24944, USA}

\author[0000-0003-0669-865X]{J.~G.~Martinez}
\affiliation{Max-Planck-Institut fur Radioastronomie MPIfR, Auf dem Hugel 69, D-53121 Bonn, Germany}

\author[0000-0001-5481-7559]{A.~E.~McEwen}
\affiliation{Center for Gravitation, Cosmology and Astrophysics, Department of Physics and Astronomy, University of Wisconsin-Milwaukee, P.O. Box 413, Milwaukee, WI 53201, USA}

\author{R.~Miller}
\affiliation{West Virginia University, Department of Physics and Astronomy, P. O. Box 6315, Morgantown, WV 26506, USA}
\affiliation{Center for Gravitational Waves and Cosmology, West Virginia University, Chestnut Ridge Research Building, Morgantown, WV 26506, USA}

\author[0000-0002-0430-6504]{E.~Parent}
\affiliation{Institute of Space Sciences (ICE, CSIC), Campus UAB, Carrer de Can Magrans s/n, 08193, Barcelona, Spain}
\affiliation{Institut d’Estudis Espacials de Catalunya (IEEC), Carrer Gran Capità 2–4, 08034 Barcelona, Spain}

\author[0000-0001-5799-9714]{S.~M.~Ransom}
\affiliation{National Radio Astronomy Observatory, 520 Edgemont Road, Charlottesville, VA 22903, USA}

\author[0000-0002-9396-9720]{M.~S.~E.~Roberts}
\affiliation{Eureka Scientific, 2452 Delmer Street, Suite 100, Oakland, CA 94602-3017, USA}

\author{A.~Rowe}
\affiliation{West Virginia University, Department of Physics and Astronomy, P. O. Box 6315, Morgantown, WV 26506, USA}
\affiliation{Center for Gravitational Waves and Cosmology, West Virginia University, Chestnut Ridge Research Building, Morgantown, WV 26506, USA}

\author{R.~Spiewak}
\affiliation{Jodrell Bank Centre for Astrophysics, School of Physics and Astronomy, University of Manchester, Manchester M13 9PL, UK}

\author[0000-0001-9784-8670]{I.~H.~Stairs}
\affiliation{Department of Physics and Astronomy, University of British Columbia, 6224 Agricultural Road, Vancouver, BC V6T 1Z1, Canada}

\author[0000-0002-7261-594X]{K.~Stovall}
\affiliation{National Radio Astronomy Observatory, 1003 Lopezville Rd., Socorro, NM 87801, USA}

\author{J.~Thorley}
\affiliation{West Virginia University, Department of Physics and Astronomy, P. O. Box 6315, Morgantown, WV 26506, USA}
\affiliation{Center for Gravitational Waves and Cosmology, West Virginia University, Chestnut Ridge Research Building, Morgantown, WV 26506, USA}

\author[0000-0001-8503-6958]{J.~van Leeuwen}
\affiliation{ASTRON, Netherlands Institute for Radio Astronomy, Oude Hoogeveensedijk 4, 7991 PD Dwingeloo, The Netherlands}

\begin{abstract}

We present the timing solutions for three radio pulsars discovered with the Green Bank North Celestial Cap (GBNCC) and 350-MHz Green Bank Telescope drift-scan surveys. These pulsars were initially discovered through their single-pulse emission and therefore designated as rotating radio transients (RRATs). Follow-up timing campaigns yielded a number of higher signal-to-noise summed pulse profiles for each pulsar, allowing us to obtain timing solutions both through single pulses as well as the standard method of time-integrating the pulsar's emission. We find that the two methods return timing parameters which are usually in agreement within two standard deviations, and have similar sized error bars. The single-pulse timing solutions have significantly higher RMS errors and reduced chi-squared values, likely due to pulse jitter. The distribution of wait times between detected single pulses indicates a significant amount of pulse clustering in time on short timescales from all three sources. For all sources, the presence of low-level emission outside of the sparse bright pulses and lack of giant pulses is more reminiscent of highly nulling canonical radio pulsars than extremely transient RRATs, highlighting the diversity of emission behavior observed from sources published as RRATs.

\end{abstract}


\section{Introduction} \label{sec:intro}
Radio pulsars are rapidly rotating, highly magnetized neutron stars with beams of radio emission extending from their magnetic poles. To date, over 4000 pulsars have been discovered (see, e.g., the ATNF pulsar catalog\footnote{v 2.6.5; \url{https://www.atnf.csiro.au/research/pulsar/psrcat/}}; \citealt{atnf}). Radio pulsars are most often discovered using Fourier-domain periodicity searches, and occasionally in the time domain using fast folding algorithms (e.g. \citealt{cbc+17,pkr+18}). 

Rotating RAdio Transients (RRATs), sources emitting isolated radio pulses separated by intervals of minutes to hours, were discovered in single-pulse searches of survey data by \citet{mll+06}. The pulses from these RRATs obey an underlying periodicity and, like standard radio pulsars, are indicative of radio emission powered by the rotational energy loss of a magnetized neutron star. 
RRATs are most commonly defined in the literature as pulsars discovered solely through their isolated pulses, as opposed to periodic emission (\citealt{mll+06,km11}). If RRATs represent a distinct and separate population of neutron stars, their sporadic nature implies the potential for an enormous population of undetected RRATs, to the point that the observed Galactic supernova rate would not be sufficient to produce the expected number of neutron stars (\citealt{mll+06,kk08}). 

The individual pulses from both pulsars and RRATs display a great deal of temporal variability in their pulse shapes and amplitudes. In addition to stochastic (random) variations of pulsed flux with time, some pulsars display sudden, discontinuous changes in their pulse shapes and fluxes. One manifestation of this phenomenon is nulling, in which the pulsar flux suddenly drops well below the detection threshold, before just as suddenly returning to its usual emission state \citep{wmj07,bb10}. Nulling pulsars display a wide range of intermittency, usually quantified by the nulling fraction, or the fraction of pulses with no detectable emission. It has been suggested that RRATs represent the extreme end of the pulsar intermittency spectrum, or manifest nulling fractions close to one \citep{bb10,fast-gpps-rrats-23,gly+25}.
However, comparisons of the RRAT and nulling pulsar populations have shown that RRATs have longer average spin periods and lower average surface magnetic field strengths; the distributions of these quantities do not appear to be drawn from the same population as those of nulling pulsars, implying that perhaps not all RRATs are just extreme nulling pulsars \citep{abhishek22}.

As more sensitive telescopes are built, it will be possible to detect low-luminosity emission outside of the bright pulses from some pulsars originally identified as RRATs, allowing for their detection in periodicity searches. Using the FAST telescope, \citet{fast-gpps-rrats-23} characterized the emission from newly- and previously-discovered RRATs, and found that they generally fall into three categories: prototypical RRATs (sporadic, isolated bright pulses separated by long spans with no detected emission), extreme nullers (several successive pulse periods with detected pulses, separated by long time spans of no detectable emission), and standard pulsars emitting low-level emission with sporadic bright pulses. Some RRATs have been observed switching between a RRAT-like and pulsar-like emission state, further indicating how observational sensitivity and the variable emission states of RRATs and pulsars can muddy the attempts to classify these objects based on long-term properties (\citealt{bb10,syw+21}). 

The single-pulse properties of RRATs can clarify their emission states and their relationship to canonical pulsars. For instance, the distribution of pulse energies of most normal pulsars and some RRATs is best described by a log-normal distribution (e.g. \citealt{sjb+12,cbm+17,mmm+18}); conversely, giant pulses with flux densities hundreds to thousands of times higher than the mean flux density have been observed from the Crab Pulsar and a handful of other pulsars, with their energy distribution following a power-law distribution (\citealt{mml+12,lmk+20,dbh+24}). Some pulsars and RRATs also show evidence of a power law tail in their energy distributions, as well as more complex pulse energy distributions (\citealt{mmm+18,fast-gpps-rrats-23}). 

The distribution of wait times between consecutive detected pulses can also be used as a diagnostic to probe the emission processes in pulsars and RRATs. If the wait times between detected pulses were governed by a random process, the probability distribution of the wait times would follow an exponential distribution with a constant burst rate, which has been observed in RRAT wait time distributions as well as giant pulses from the Crab pulsar (\citealt{kss+10,bsa+18}). Some RRATs show evidence of pulse clustering on short timescales, with their wait time distributions exhibiting non-Poissonian behavior (\citealt{bsa+18,bbc+22,gly+25}). \citet{oppermann+18} showed that a Weibull distribution with a clustering parameter $k$ better fits the pulse arrival times of extragalactic fast radio bursts; $k<1$ means pulses are more likely to be closely spaced than in the simple Poissonian case.

The rotational period and period derivative, sky coordinates, and other astrometric parameters of a pulsar or RRAT are constrained through the process of pulsar timing (for a more in-depth review, see \citet{hbopa}). The arrival times of pulses are measured and iteratively fit to a rotational model of the pulsar through least-squares minimization of the residuals, or differences between the observed and predicted arrival times. Pulsar timing accounts for every rotation of the pulsar over the span of the solution, allowing for the measurement of the neutron star's rotational and astrometric parameters with extreme precision. These parameters can then be used to estimate canonical properties of the neutron star such as the surface magnetic field strength and spin-down luminosity. Deriving RRAT timing solutions is therefore crucial for estimating their age and magnetic field strength, and characterizing their evolutionary history.

Most pulsars are weak radio emitters, and their individual pulses are not sufficiently brighter than the background noise of the receiver system to allow timing of single pulses. Instead, hundreds to thousands of pulses are added \textit{modulo} the pulse period to form an integrated pulse profile. This has the additional benefit of being more stable in time due to the reduction of intrinsic pulse-to-pulse jitter, or random variations in pulse phase and amplitude \citep{lma+19}. In the case of the most sporadic RRATs, however, there are typically not enough detected pulses to add up to create a sufficiently high signal-to-noise summed profile, and timing must be performed through measurements of the single-pulse arrival times \citep{mlk+09}.

In this paper, we present the timing solutions of three pulsars which were initially discovered only through their RRAT-like single pulses, but which were detectable through periodicity searches at a later date.  We compare the efficacy of timing through both integrated profiles and single pulses, and use the single-pulse statistics to draw conclusions about the emission mechanisms at play. We describe the observations in Section~\ref{sec:obs}, and both forms of TOA generation and timing analysis in Section~\ref{sec:timing}. In Section~\ref{sec:sps} we describe the single-pulse analysis including the wait time and pulse energy distributions. We discuss our findings in Section~\ref{sec:disc} and conclude in Section~\ref{sec:con}.

\section{Observations and Data Reduction}
\label{sec:obs}

\input{observations}

\begin{figure}
    \centering
    \includegraphics[width=\linewidth]{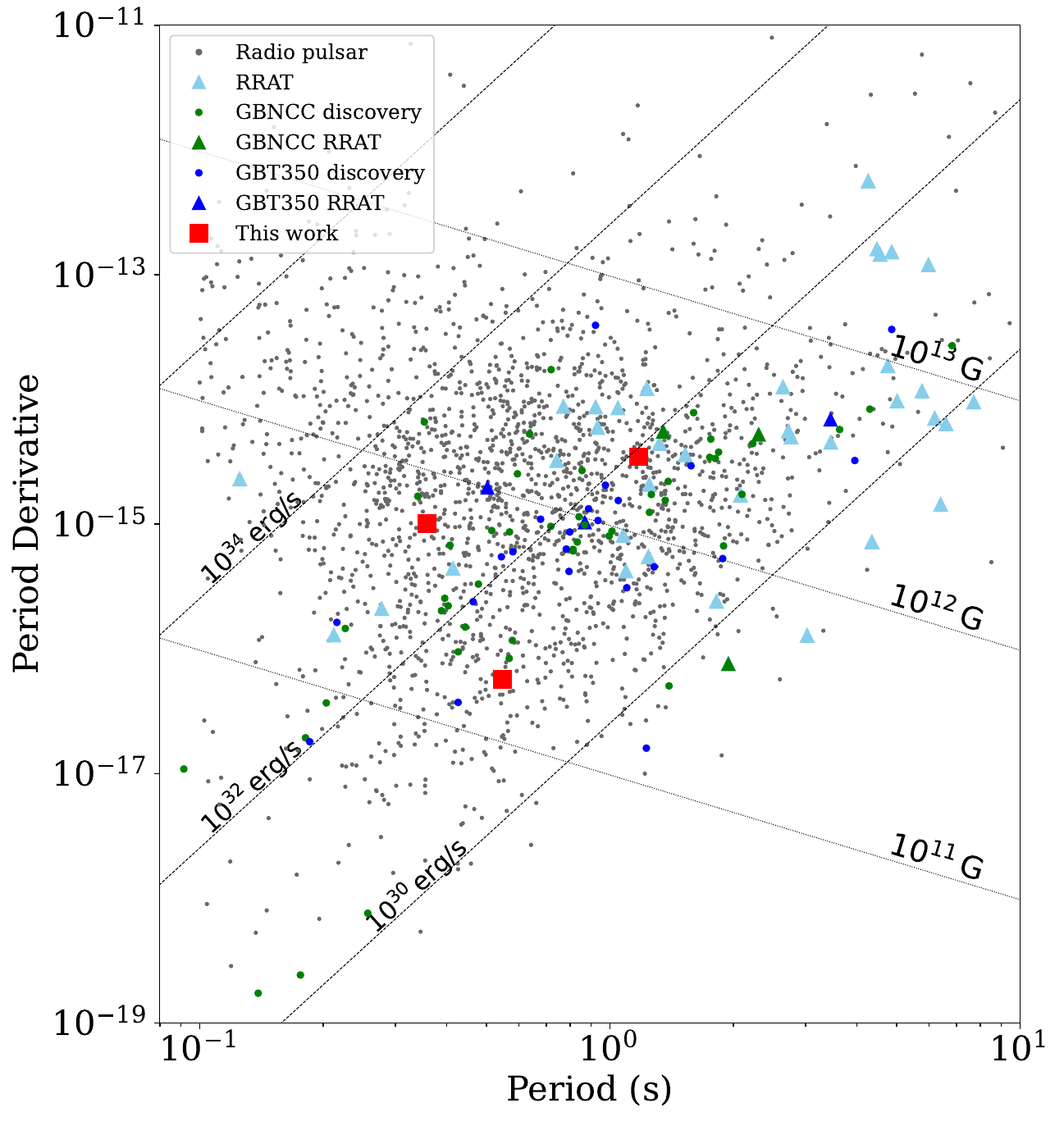}
    \caption{$P-\dot{P}$ diagram showing the population of radio pulsars as a function of rotational period and period derivative. The three sources studied in this work are denoted with red squares. Pulsars and RRATs discovered by the GBNCC survey are labeled in green, and those discovered by the GBT350 survey in blue. Lines of constant spin-down energy and constant surface magnetic field are labeled.}
    \label{fig:PPdot}
\end{figure}

\begin{figure}
    \centering
    \includegraphics[width=\linewidth]{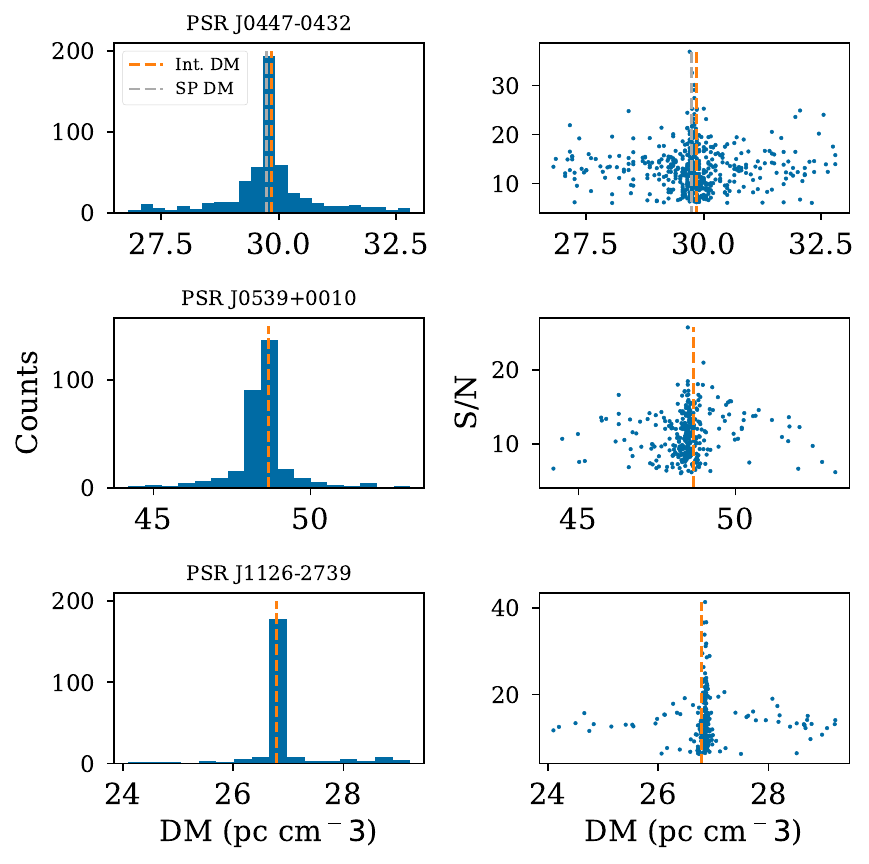}
    \caption{ The left column contains a histogram of the best-fit DMs for each single pulse, and the right column shows a scatter plot of the best-fit DM vs. the S/N of that pulse reported by \textsc{presto}. For PSR J0447--0432, the orange vertical dashed line represents the DM obtained through timing with integrated TOAs, while the gray dashed line represents the DM obtained through timing with single-pulse TOAs. For the other two pulsars, the pulsar DM was not constrained through timing, and the fiducial DM is shown with an orange vertical dashed line.}
    \label{fig:DM_dists}
\end{figure}

\begin{figure}
    \centering
    \includegraphics[width=0.9\linewidth]{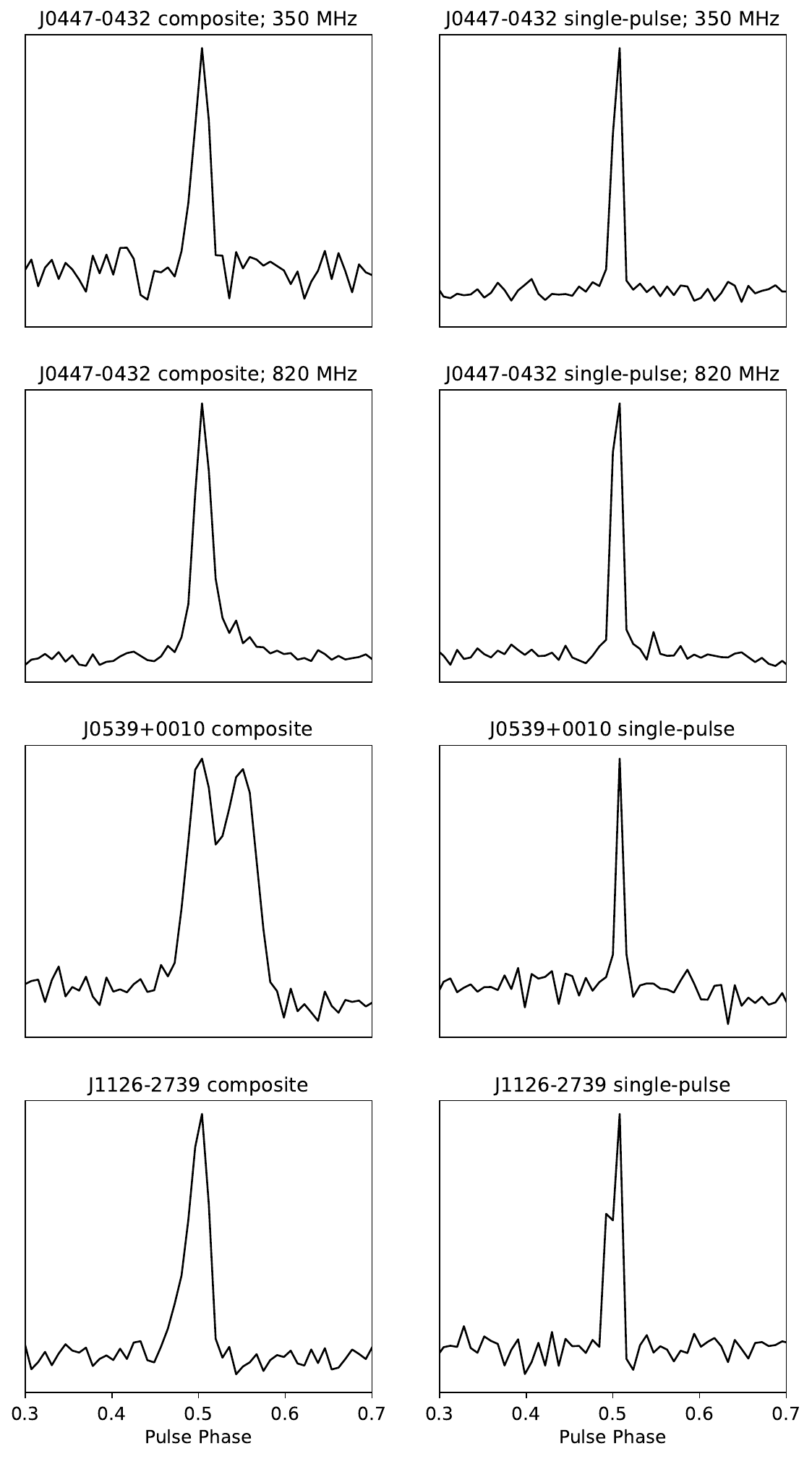}
    \caption{Left: integrated pulse profiles for each pulsar. Right: brightest single pulse for each pulsar. Each profile covers 40\% of a single pulsar rotation. The vertical axes are in units of arbitrary intensity.}
    \label{fig:profs}
\end{figure}

\begin{figure}
    \centering
    \includegraphics[width=0.85\linewidth]{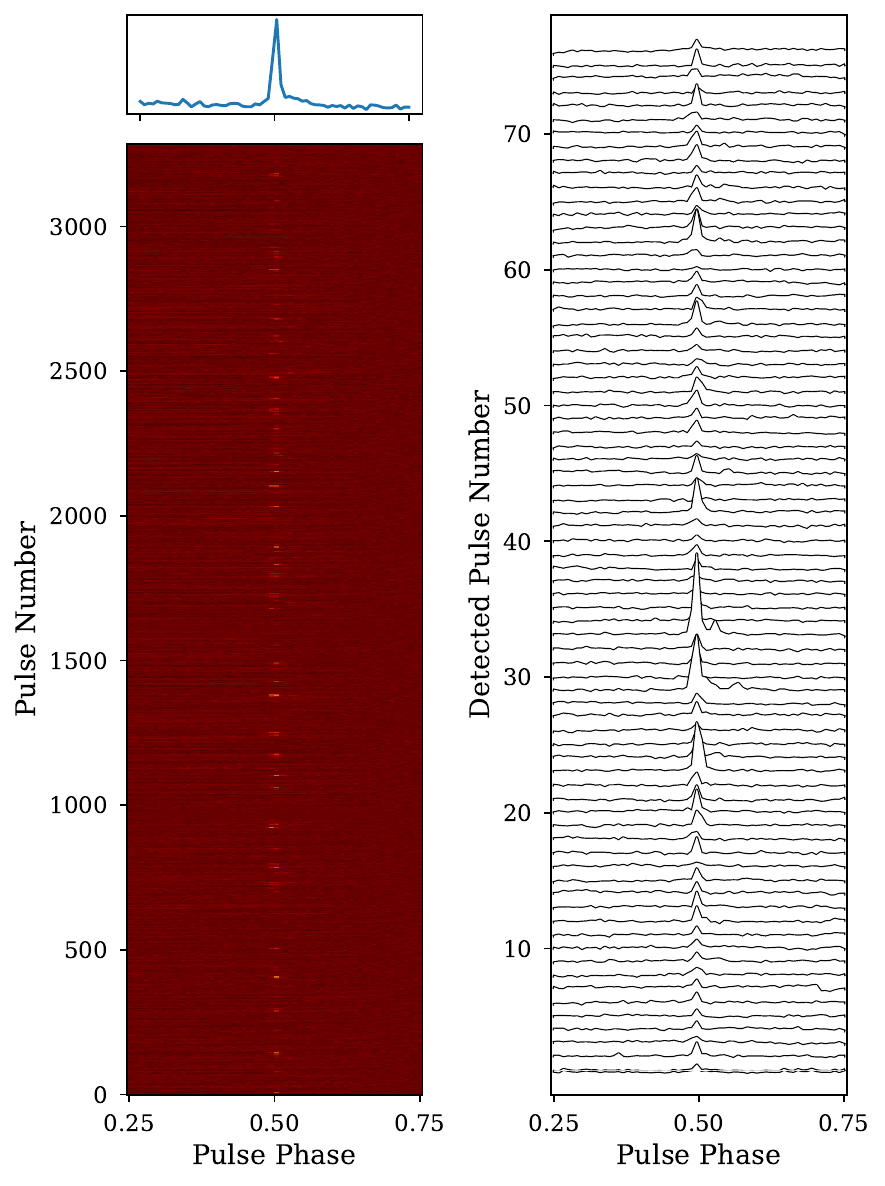}
    \caption{Emission plots for one observation of PSR J0447--0432. Left: a time-phase plot of all 3286 rotation periods in the observation, with the integrated pulse profile above. Right: a sequential pulse stack of the profiles of 75 single pulses from this observation with measured times of arrival in the timing solution.}
    \label{fig:0447_timephase}
\end{figure}

The three pulsars studied in this paper were discovered through their single-pulse emission in survey observations from the 350-MHz Green Bank Telescope (GBT) drift-scan survey (GBT350; \citealt{blr+13,lbr+13}) and the the Green Bank North Celestial Cap (GBNCC) survey \citep{slr+14,kmk+18,lsk+18,acd+19,mss+20}. 

The GBT350 drift-scan survey has discovered 35 pulsars\footnote{\url{https://astro.phys.wvu.edu/GBTdrift350/}}, indicated in dark blue in the $P-\dot{P}$ diagram shown in Figure~\ref{fig:PPdot}. The GBNCC scanned the entire Northern sky visible to the GBT and  has discovered 195 pulsars including 24 RRATs\footnote{\url{https://astro.phys.wvu.edu/GBNCC/}}; GBNCC discoveries are highlighted in green in Figure~\ref{fig:PPdot}. 

PSRs J0447--0432 (hereafter J0447) and J0539+0010 (hereafter J0539) were discovered in GBT350 data, although J0539 was never published. The sky position and dispersion measure (DM) of J0447 were reported in \citet{KA15}, as well as an estimated rotational period determined through the arrival time differences between detected single pulses. J0539 was also detected by the Commensal Radio Astronomy FAST Survey (CRAFTS\footnote{\url{http://crafts.bao.ac.cn/}}; \citealt{crafts}) and labeled on their discovery webpage as PSR J0539+0013, although not published. 

PSR J1126--2739 (hereafter J1126) was discovered\footnote{\url{http://astro.phys.wvu.edu/GBNCC/J1126-27_sp.png}} by the GBNCC survey, and identified using the \texttt{RRATTrap} single-pulse sifting algorithm\footnote{\url{https://github.com/ckarako/rrattrap}} \citep{KA15}. 

Timing follow-up observations\footnote{Observing programs: AGBT14A\_507, AGBT15A\_376, AGBT16A\_343, AGBT18B\_361, AGBT19A\_180, AGBT19B\_306, and AGBT19B\_320} were undertaken using the GUPPI backend \citep{guppi} on the GBT at 350 and 820 MHz; details are available in Table~\ref{tab:obs}.

Using the \textsc{presto}\footnote{\url{https://github.com/scottransom/presto}} \citep{presto} software package, we identified and masked impulsive RFI signals in our data using \texttt{rfifind}. In addition, for observations at 350 MHz, persistent RFI occurs in the 360--380 MHz range, and the corresponding channels were ignored in all subsequent analyses.

\subsection{TOA Generation-- Single Pulses}
As pulsar signals travel through ionized material in the interstellar medium, the signal incurs a frequency-dependent delay proportional to the amount of free electrons along the observer's line of sight, with the constant of proportionality known as the dispersion measure (DM). A single broadband pulse from a pulsar arrives later at lower observing frequencies compared to higher frequencies, and the data must be dedispersed in order to account for this effect and maximize the detectability of the pulsar signal.

We used \textsc{presto}'s \texttt{prepdata} function to create a time series from the RFI-mitigated data at a DM of 0 pc~cm$^{-3}$, and \texttt{prepsubband} to create many de-dispersed time series over a range of DMs centered on the known DM of the pulsar. 
The DM spacing between time series and the downsampling factor used were calculated  using \textsc{presto}'s \texttt{DDplan.py}, and are given in Table~\ref{tab:obs}. We then used \texttt{single\_pulse\_search.py} to search each time series for pulses, recording any candidate single-pulse events with signal-to-noise ratios (S/Ns) greater than 5. We use an expanded list of boxcar widths in order to be sensitive to a variety of burst widths, and use a maximum boxcar width which corresponds to a pulse width of 100 ms.

We search over a range of DMs to discriminate astrophysical faint pulses from  RFI, as
the detected S/N of a single pulse from a pulsar will be decreased if the data are de-dispersed at an incorrect DM \citep{cm03}. The S/N is expected to be maximized at the DM of the pulsar, and fall off on either side; thus, even if the single-pulse search code repeatedly identifies the same pulse in multiple de-dispersed time series, the grouping of single-pulse events will be closely spaced in time, and the pulse can be identified by choosing the event with the highest S/N. 

We observe variations in the DM that maximizes the S/N between single pulses, likely due to radiometer noise, as is evident by the distribution of single-pulse DMs in Figure~\ref{fig:DM_dists}. Thus, to avoid discarding real pulses, we accept single pulses with peak S/Ns at DMs within 10\% of the fiducial DM, which is either determined by the peak DM of the brightest astrophysical pulse detected, or the DM measured through timing with integrated pulse profiles. In order to mitigate RFI which may be falsely flagged as a real pulse, a candidate is removed if there is a pulse at a DM of 0~pc~cm$^{-3}$ with a greater S/N within one pulse period of the candidate. 

We created a template profile at each observing frequency by fitting a Gaussian template to the brightest single pulse, and measured a TOA for each pulse by cross-correlating it with this template profile. 
The errors on the TOA are approximated as the width of a bin divided by the signal-to-noise ratio of the pulse, as in \citet{mlk+09}.
We visually inspected each pulse to confirm their astrophysical nature, removing pulses with baselines greatly affected by RFI.

\subsection{TOA Generation-- Integrated emission}
These pulsars were all discovered through their single pulses, but in later follow-up observations, were sufficiently luminous to be detected by an FFT search at several observational epochs. In \citet{KA15}, the rotational period of J1126 was  constrained using the time differences between detected single pulses, and the quoted rotational period of J0447 was a factor of four longer than the period we report here. 

We used the FFT detections as the nominal rotational periods for each pulsar. After folding each observation, we combined the profiles for each detection into a high signal-to-noise (S/N) composite profile; the profiles for each pulsar are shown in the left panel of Figure~\ref{fig:profs}. We then generated TOAs for each observation in which each pulsar was detected using the template profile and \textsc{PRESTO}'s \texttt{get\_TOAs.py}; the TOA uncertainties were also reported by \textsc{PRESTO}. We generally used only one TOA for each observation except for a `phase-connection' observation, an observation with a high-S/N detection which is closely spaced in time to other observations. An exemplary time-phase plot, integrated pulse profile, and set of single-pulse profiles are presented in Figure~\ref{fig:0447_timephase}. 

\section{Timing Solutions}
\label{sec:timing}

We used the TEMPO2 software package to obtain all of the timing solutions, using the DE438 solar system ephemeris. We multiplied the uncertainty on each TOA so that the reduced $\chi^2$ value of the post-fit timing solution would be equal to one. 

Since we were able to obtain TOAs at both 350 and 820 MHz for J0447, we also fit for DM in the process of pulsar timing, and the reported DM and error are obtained from the timing solution. For J0539 and J1126, we were only able to obtain TOAs at 350 MHz, and thus could not refine the DM measurements through timing. We attempted to fit for the DM by splitting the observing band into subbands and measuring a TOA within each subband, however, performing this analysis on the highest S/N observations for each pulsar did not yield more precise DM measurements than those previously constrained. Instead, we selected the 20 brightest pulses from our sample and took the weighted average of their PRESTO-reported DM, assuming a 5\% uncertainty on the DM reported in the sections below.

\subsection{PSR J0447--0432}
\begin{figure}
    \centering
    \includegraphics[width=\linewidth]{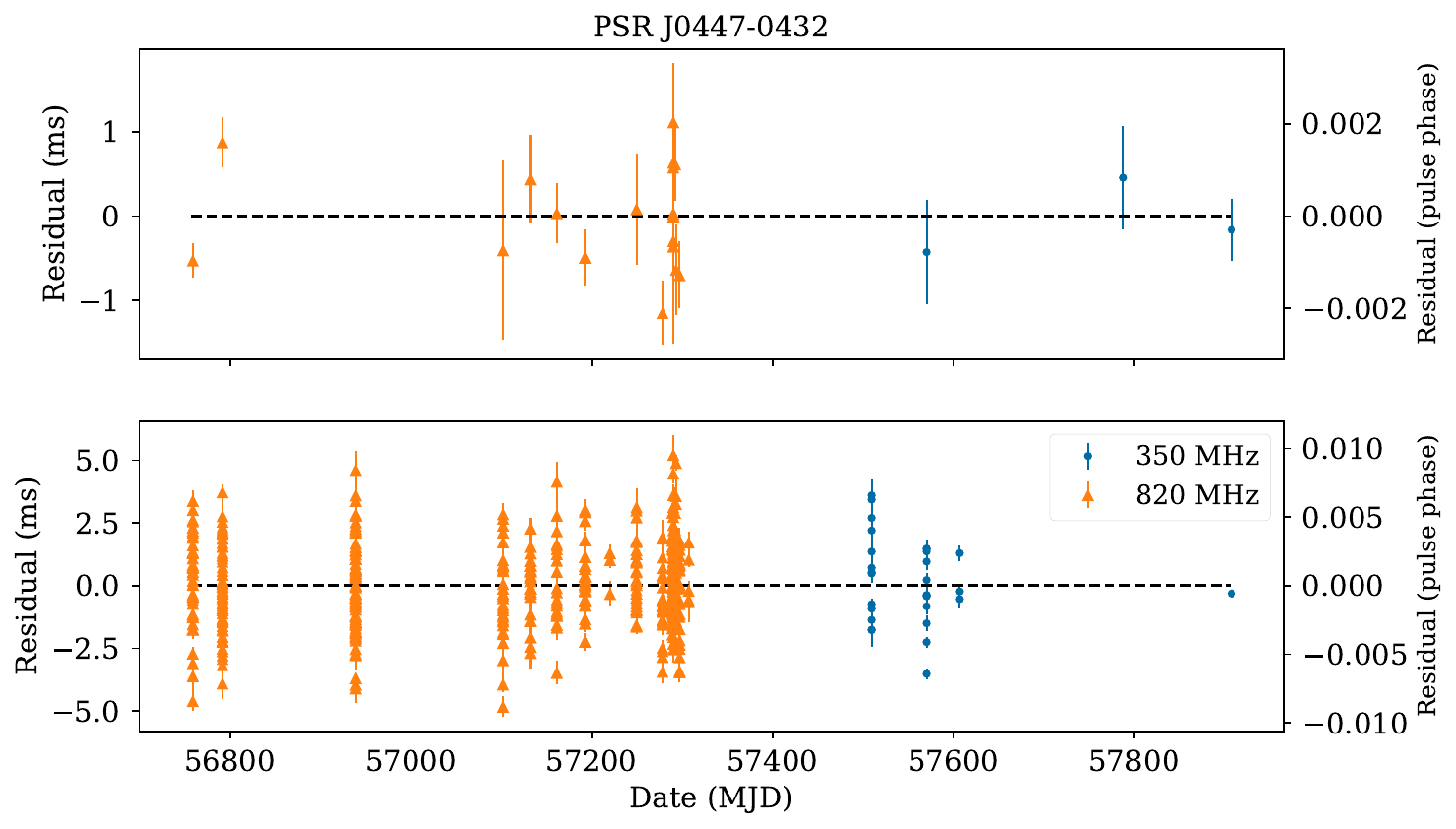}
    \caption{Top: the timing residuals for J0447 using integrated timing.  Orange triangles represent the residuals for TOAs at 820 MHz, and blue circles the residuals at 350 MHz. Bottom: the timing residuals for J0447 using single-pulse timing, with the same color mapping.}
    \label{fig:0447resids}
\end{figure}

The full timing solution for J0447 is given in Table~\ref{tab:0447}, and the timing residuals are plotted in Figure~\ref{fig:0447resids}. The derived sky positions and rotational parameters of the two methods are in agreement within two standard deviations. 

The rotational period of J0447 was estimated through the single pulse arrival times in \citet{KA15}. According to the reported burst rate estimates, no more than three pulses would have been detected in a given observation. The use of the greatest common denominator between pulse times as an estimate for a RRAT's rotational period can often lead to a period estimate greater than the true rotational period, and simulations by \citet{cbm+17} show that the probability of determining an incorrect period from three pulses in an observation is roughly 37\%; thus, the overestimated period for J0447 is not surprising. 

There are five epochs with single pulse TOAs but no integrated TOA in the timing solutions, and one epoch with an integrated TOA but no single pulse TOAs. Three of the epochs with only single pulse TOAs were significantly corrupted by RFI so that an integrated TOA was not obtainable, and the other two observations were shorter than ten minutes long and yielded a signal that was too weak to reliably measure an integrated TOA. The observation with an integrated TOA but no single pulse TOAs was at 350 MHz and there were no single pulses with S/N $>$ 5 detected by \textsc{PRESTO}; the lack of bright single pulses at 350 MHz is discussed further in Section~\ref{sec:pulserates}.

\input{0447table}

\subsection{PSR J0539+0010}
\label{sec:0539}

\begin{figure}
    \centering
    \includegraphics[width=\linewidth]{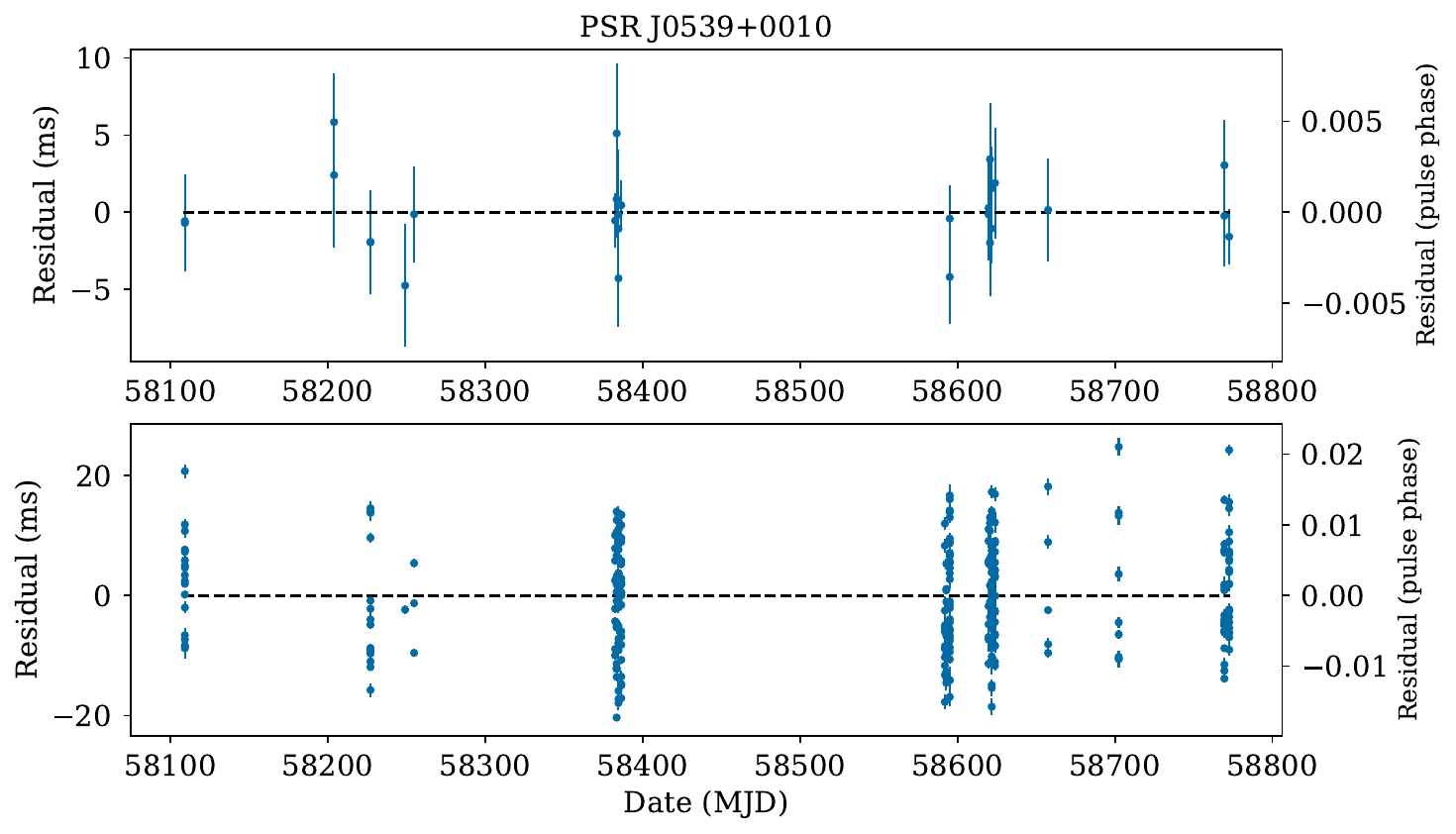}
    \caption{Top: the timing residuals for J0539 using integrated timing. All TOAs have been measured at 350 MHz. Bottom: the timing residuals for J0539 using single-pulse timing. The single-pulse TOAs have been adjusted to a single band, as described in Section~\ref{sec:0539}.}
    \label{fig:0539resids}
\end{figure}

\begin{figure}
    \centering
    \includegraphics[width=\linewidth]{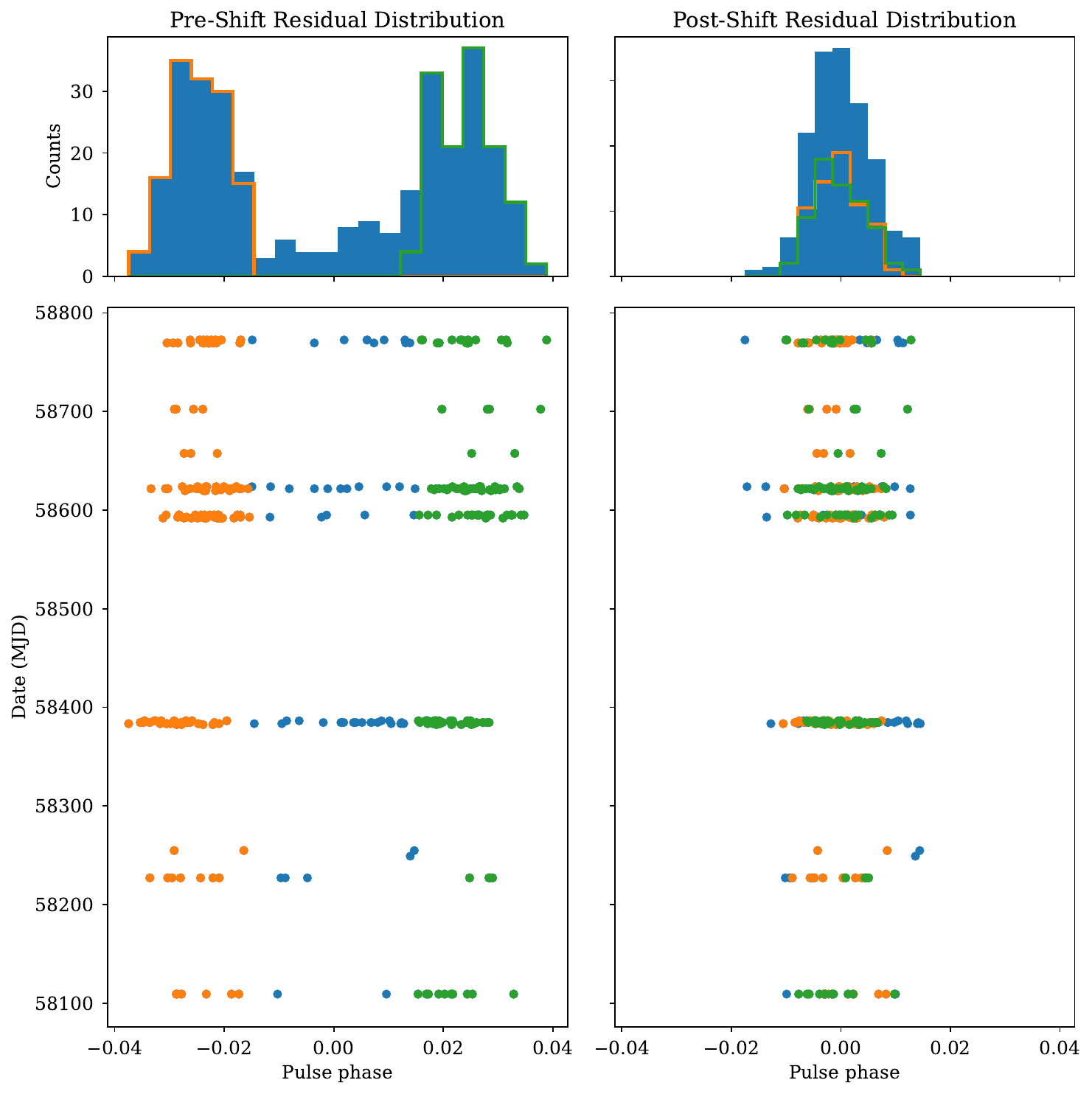}
    \caption{Left: the distribution of timing residuals from the TOAs of the single pulses from PSR J0539+0010 over the 1.8--year observational campaign, with the residuals as a function of MJD plotted below. The TOAs tend to cluster in two peaks, denoted by orange and green, respectively. Right: the distribution of timing residuals from the same TOAs, after adjusting the TOAs as described in the text. The TOAs originally in the early or late band are still denoted by orange and green, respectively.}
    \label{fig:0539_resid_shift}
\end{figure}

The full timing solution for J0539 is given in Table~\ref{tab:0539}, and the timing residuals are plotted in Figure~\ref{fig:0539resids}. The derived  rotational parameters of the two methods are in agreement within two standard deviations.


The integrated profile of J0539 consistently shows a double-peaked shape across observational epochs. The residuals of the single-pulse TOAs cluster within two main bands, with approximately 17\% of pulse residuals falling between the two peaks, and the remainder distributed equally across both bands. The left panel of Figure~\ref{fig:0539_resid_shift} shows the distribution of the residuals of the single-pulse TOAs, using the best-fit timing solution derived from these TOAs. The bands are centered roughly at pulse phases of $\pm$0.025, equivalent to $\pm 29.4$ ms from the center of the profile. 

Pulsar timing methodology is based on the assumption that the pulse originates from the  same rotational phase every time. In order to obtain a single-pulse timing solution with lower uncertainties, we first wanted to find the optimal amount by which to shift the TOAs in each band to align the TOAs to a single peak, as in \citet{lmk+09}. We defined the early and late bands as being composed of TOAs with residuals with absolute values $> 0.015$ in phase, and searched 30 offsets between pulse phases of 0.02 and 0.03 for each band, creating a set of TOAs with the early- and late-band TOAs shifted by the corresponding amount. We chose the offsets corresponding to the set of TOAs which, when run through \textsc{tempo2}, yielded the timing solution with the lowest RMS residual. This resulted in offsets of 34.9 ms for the early band and 32.5 ms for the late band. The RMS residual on the solution decreased from 27.2 ms to 7.2 ms after the TOAs were adjusted, and the uncertainties on the parameters by similar factors.

There are three RFI-impacted epochs with single pulse TOAs but no integrated TOA in the timing solutions, and there is one epoch with an integrated TOA but no single pulse TOAs, as no single pulses with S/N $>$ 7 were detected by PRESTO, and the few single pulses with 5 $<$ S/N $<$ 7 were not suitably bright for a reliable TOA measurement.

\input{0539table}

\subsection{PSR J1126--2739}

\begin{figure}
    \centering
    \includegraphics[width=\linewidth]{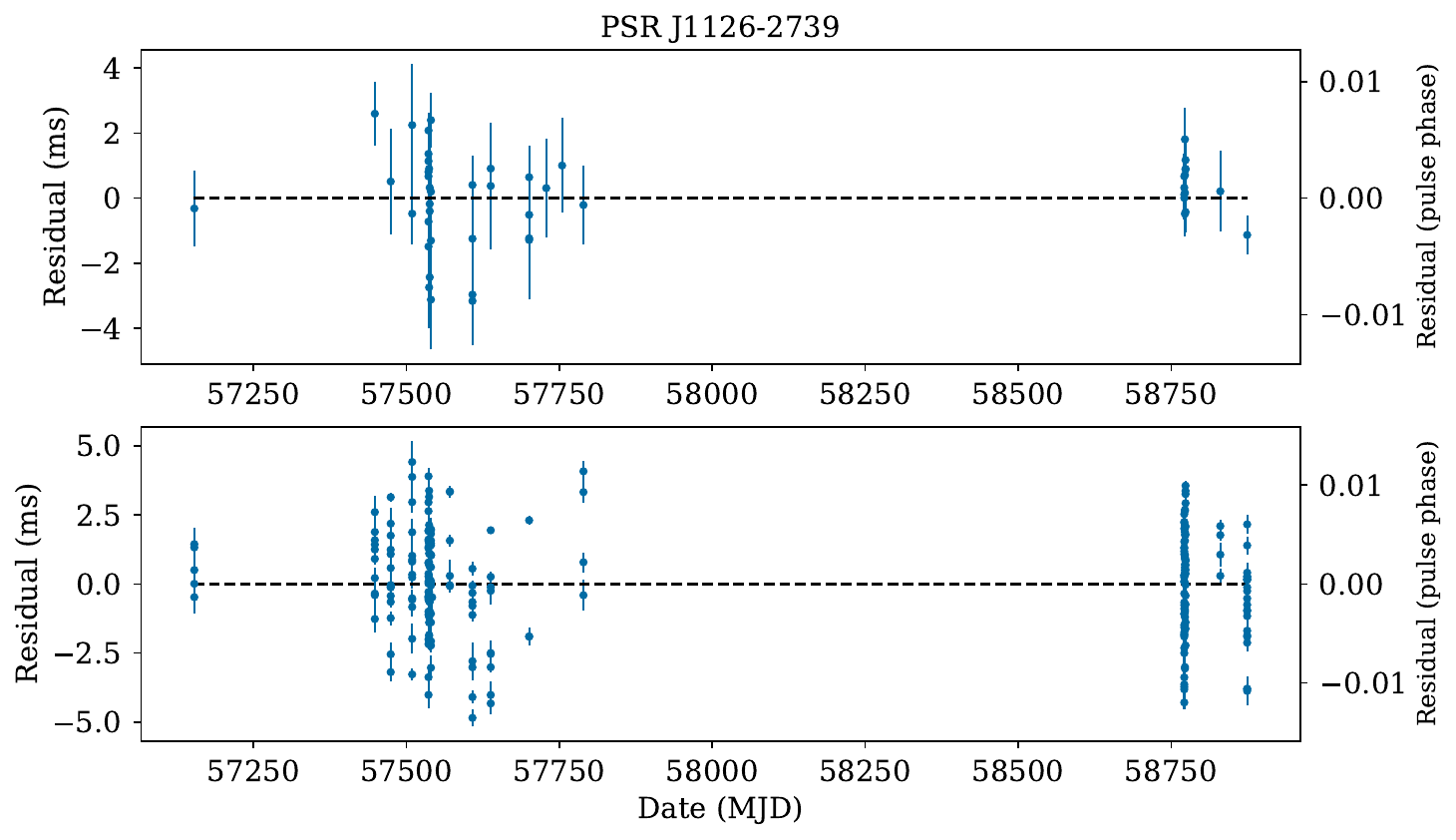}
    \caption{Top: the timing residuals for J1126 using integrated timing. All TOAs have been measured at 350 MHz. Bottom: the timing residuals for J1126 using single-pulse timing.}
    \label{fig:1126resids}
\end{figure}

The full timing solution for J1126 is given in Table~\ref{tab:1126}, and the timing residuals are plotted in Figure~\ref{fig:1126resids}. The  rotational parameters of the two methods are in agreement within two standard deviations, and the sky positions are in agreement within 2.5 standard deviations.

There are two RFI-impacted epochs with single pulse TOAs but no integrated TOA in the timing solutions, and two epochs with an integrated TOA but no single pulse TOAs, as no single pulses with S/N $>$ 7 were detected by PRESTO, and the few single pulses with 5 $<$ S/N $<$ 7 were not suitably bright for a reliable TOA measurement.

\input{1126table} 

\section{Emission Properties}
\label{sec:sps}
\subsection{Pulse Rates}
\label{sec:pulserates}

\begin{figure}
    \centering
    \includegraphics[width=\linewidth]{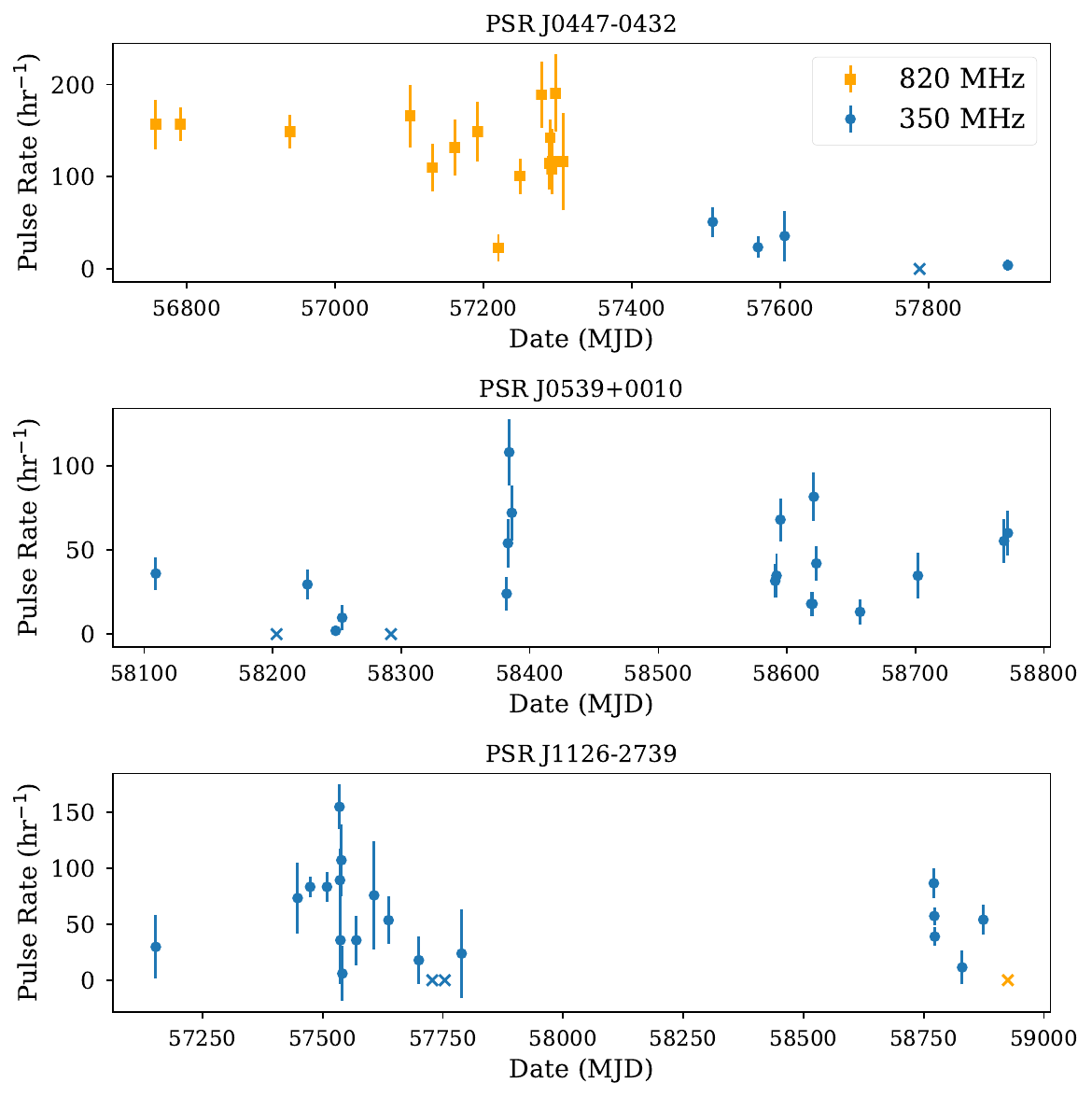}
    \caption{The detected pulses rate for each observation of PSRs J0447--0432 (top), J0539+0010 (middle), and J1126--2739 (bottom). Observations at 820 MHz are marked with orange squares, and observations at 350 MHz with blue circles. Observations with no detected emission are marked with an ``x" in the color corresponding to the central frequency of the observation. The error bars correspond to 1-$\sigma$ error bars corresponding to Poissonian statistics \citep{g86}.}
    \label{fig:burstrate}
\end{figure}

The rate of detected pulses within each observation, defined as the number of single pulse TOAs divided by the observation length, is shown for all three pulsars in Figure~\ref{fig:burstrate}. J0447 has consistently higher pulse rates at 820 MHz (mean rate of 132 pulses hr$^{-1}$) compared to 350 MHz (mean rate of 28 pulses hr$^{1}$), which is also clear from the single-pulse residuals in Figure~\ref{fig:0447resids}. The predicted scattering timescales are roughly 0.4 and 730 $\mu$s at 820 and 350 MHz respectively, so the signal is not scattered out at lower frequencies, and we do not see strong scattering tails in the profiles of this pulsar at 350 MHz (e.g. Figure~\ref{fig:profs}). 

At 350 MHz, the system temperature is higher, and the effective bandwidth is roughly one third of that of the 820 MHz receiver. Using the radiometer equation, a detection threshold S/N of 6 implies a single pulse flux detection threshold of 8 mJy at 820 MHz, and 20 mJy at 350 MHz. We also used the radiometer equation to estimate the mean flux of J0447 in each observation, finding an average flux of $56 \pm 35$ mJy at 820 MHz and $49 \pm 21$ mJy at 350 MHz, where the error bars are the standard deviations of all of the flux measurements at that frequency.

Assuming that J0447 follows a simple spectral power law with the average pulsar spectral index of $-1.6$ \citep{jvk+18}, the mean flux at 350 MHz would be roughly four times higher than that at 820 MHz, and we would expect more single pulses exceeding the detection threshold at 350 MHz than at 820. Conversely, the detected pulse rate at 820 MHz is on average 4.5 times greater than the pulse rate at 350 MHz, though the time-averaged fluxes are very similar between the two frequencies. We thus conclude that there are fewer, though brighter, detected single pulses at 350 MHz compared to at 820 MHz where the single-pulse detection threshold is lower.

\subsection{Wait Times and Pulse Clustering}
\begin{figure*}
    \centering
    \includegraphics[width=\linewidth]{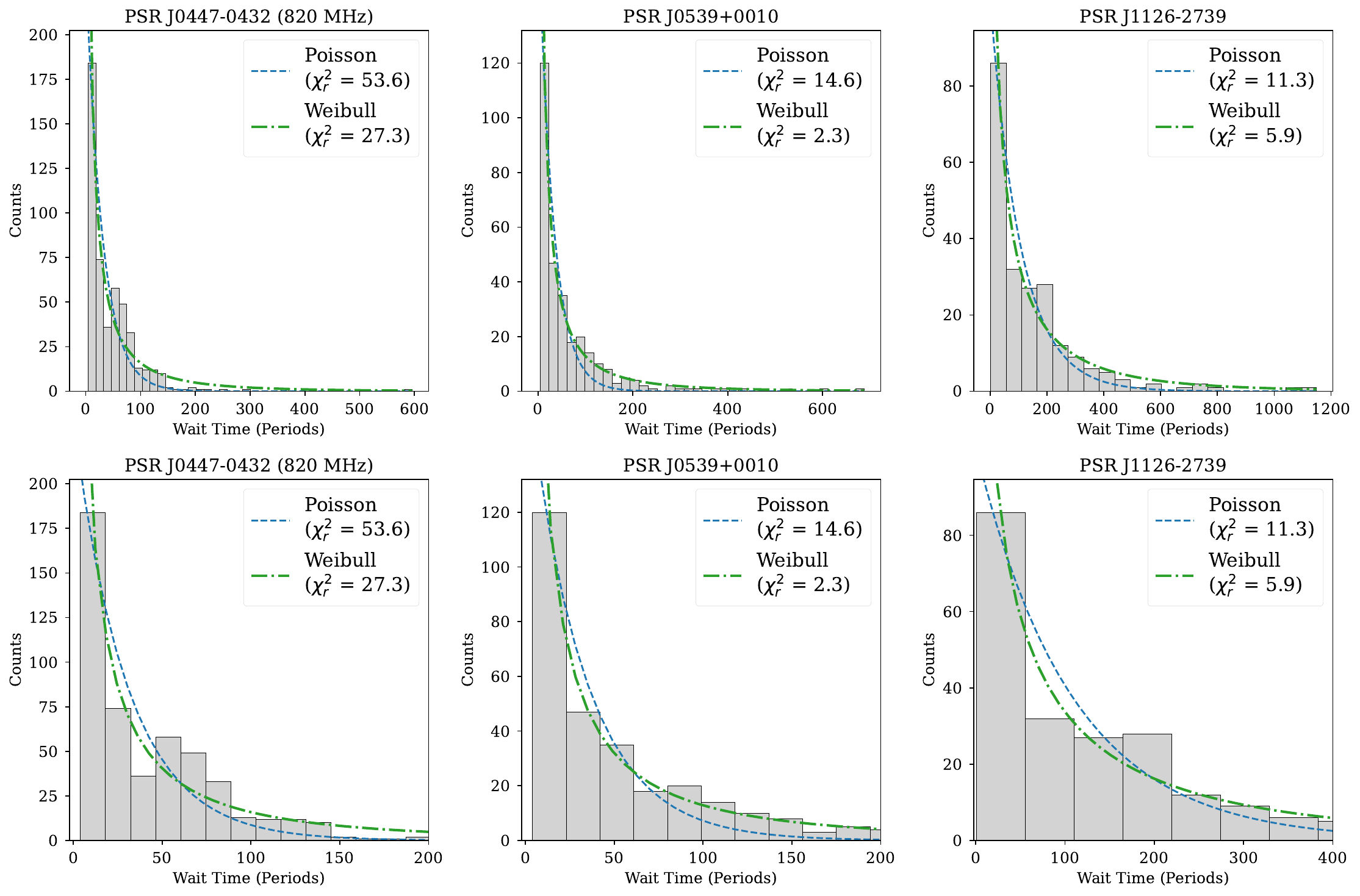}
    \caption{Wait time distributions for all three pulsars studied in this paper. The best-fit Poisson distribution is overlaid with a blue dashed line, and the best-fit Weibull distribution in a dot-dashed green line. The bottom row displays the same distributions and analytical fits as the top row, zoomed in to show only the first 200 or 400 periods.}
    \label{fig:wait_time}
\end{figure*}

\input{waittime_dists}

We used the measured arrival times of each pulse to create a wait time distribution, excluding any wait times between observations due to the sporadic cadence of our observations. We fit Poisson and Weibull distributions to each distribution using \texttt{scipy.optimize.curve\_fit} and report the results of each fit in Table~\ref{tab:wts}. The number of bins in each distribution were chosen using the Freedman Diaconis estimator. The wait times, in units of pulsar rotations, are shown in Figure~\ref{fig:wait_time} along with the fitted statistical distributions. 

The distribution of wait times from J0447 shows slight evidence of a secondary peak of wait times centered at around 75 periods, although the inclusion of a Maxwell-Boltzmann or Gaussian component did not yield a statistically significant improvement in the quality of the fit. Similar secondary `bumps' in the wait time distribution were observed in PSRs J1317--5759 and J1819--1458, with increased wait times at $\sim$ 25 pulse periods, although no statistical distribution was able to fully describe the bumps in the distribution, likely due to the complex nature of the wait time distributions \citep{bsa+18}. 

Since only the TOAs from the timing analysis were used to calculate the wait times, low-luminosity pulses would not be considered, potentially introducing a bias towards longer wait times. Additionally, excising rotations corrupted by RFI could cause astrophysical pulses to be excised, artificially increasing the reported wait times. To measure the effect of the long-wait tails on the distribution fits, we re-ran the wait time analysis only on wait times shorter than a given threshold; cutoff thresholds of 200, 250, and 500 rotational periods each yielded results in agreement with those reported in Table~\ref{tab:wts}. Overall, we find that the Weibull distribution with $k < 1$ is preferred for all three pulsars, indicating that pulse clustering on short timescales is present. 

We performed a statistical F-test to compare the goodness of fit of the two distributions, and determine if adding the extra parameter in fitting the Weibull distribution makes a statistically significant improvement to the fit. The F-test returns a statistic which we use to determine the probability that the improvement to the fit is due to random chance. These false alarm probabilities are presented in the final column of Table~\ref{tab:wts}, and indicate that the wait times from J0539 are much better fit by a Weibull distribution, and the same is true to a lesser extent for J0447. We conclude that significant amounts of pulse clustering are present in the emission from PSRs J0447 and J0539, and that the pulse clustering model may be preferred for PSR J1126, but with lower certainty.


\subsection{Pulse Energy Distributions}

\begin{figure}
    \centering
    \includegraphics[width=\linewidth]{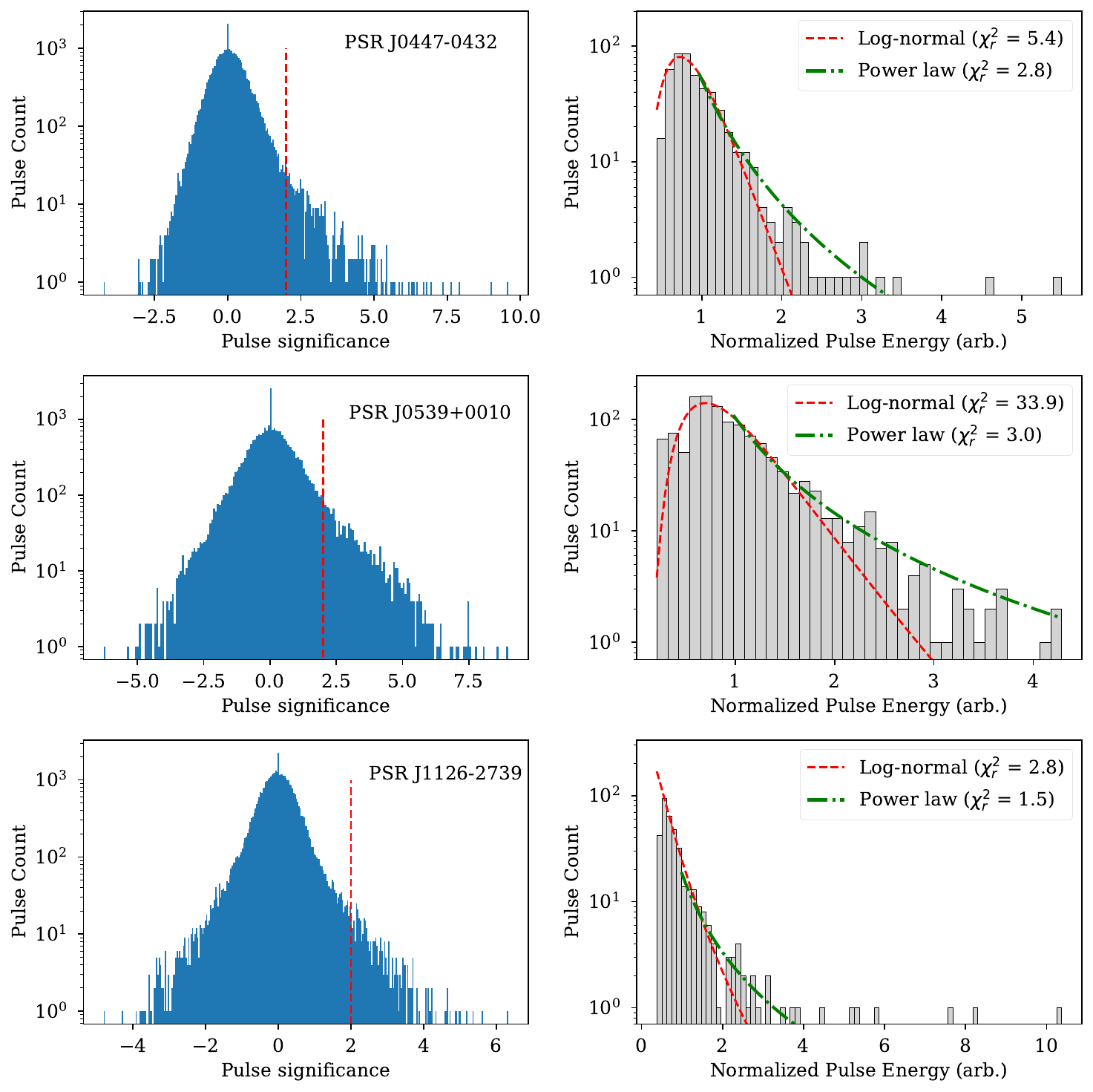}
    \caption{The pulse energy distributions for all three pulsars studied in this paper. Left: the distribution of on-pulse S/N for every rotation observed from each pulsar. The red dashed lines denote cutoffs of S/N = 2. Narrow peaks at S/N = 0 correspond to RFI-corrupted data which have been masked. Right: the distribution of pulse energies for all pulses with S/N $\geq 2$.  The best-fit log-normal distribution is overlaid with a red dashed line, and the best-fit power law distribution in a dot-dashed green line.}
    \label{fig:energy_dists}
\end{figure}

We determined on-pulse and off-pulse regions for each pulsar by visible inspection of their time-phase plots. For each pulse, the on-pulse energy and off-pulse standard deviation were calculated, and normalized by the average on-pulse energy of all rotations from that observation. 
The resulting distributions are shown in Figure~\ref{fig:energy_dists} (left panel). 
In nulling pulsars of sufficient brightness such a distribution naturally separates into two distinct components \citep[see, e.g.,][]{lkr+02}, 
allowing an unambiguous distinction between pulses with and without emission. 
In these three cases, however, the energy distributions are continuous. 
We choose $\mathrm{S/N > 2}$  as the boundary, and study the pulse energy distribution of pulses exceeding that. 
The pulse energies of all these pulses with $\mathrm{S/N > 2}$ are next plotted for each pulsar in the right panel of Figure~\ref{fig:energy_dists}, alongside the best-fit log-normal and power law distributions fit using \texttt{scipy.optimize.curve\_fit}. A combined log-normal and power-law distribution was also tested, but led to worse fits for all three pulsars, so we do not plot them. 

The fit results indicate that a power-law distribution is preferred over a log-normal energy distribution for all three pulsars in this sample. A power law tail in the energy distribution implies the presence of giant pulses; however, giant pulses from the Crab pulsar and others are several orders of magnitude brighter than the average pulsed flux. We do not see such a pronounced increase in pulse energy in the brightest pulses from this sample, with only one pulse from J1126 being greater than ten times more energetic than the average pulse energy. The lack of exceptionally bright pulses leads us to conclude that the emission mechanisms are more similar to the standard pulsar model, as opposed to giant pulse emission, despite the slight preference for power law distributions. 
We also examined the pulse energies as a function of the wait time between detected pulses finding no correlation, suggesting that energy is not being stored up between pulses.

\subsection{Weak Emission}

\begin{figure}
    \centering
    \includegraphics[width=\linewidth]{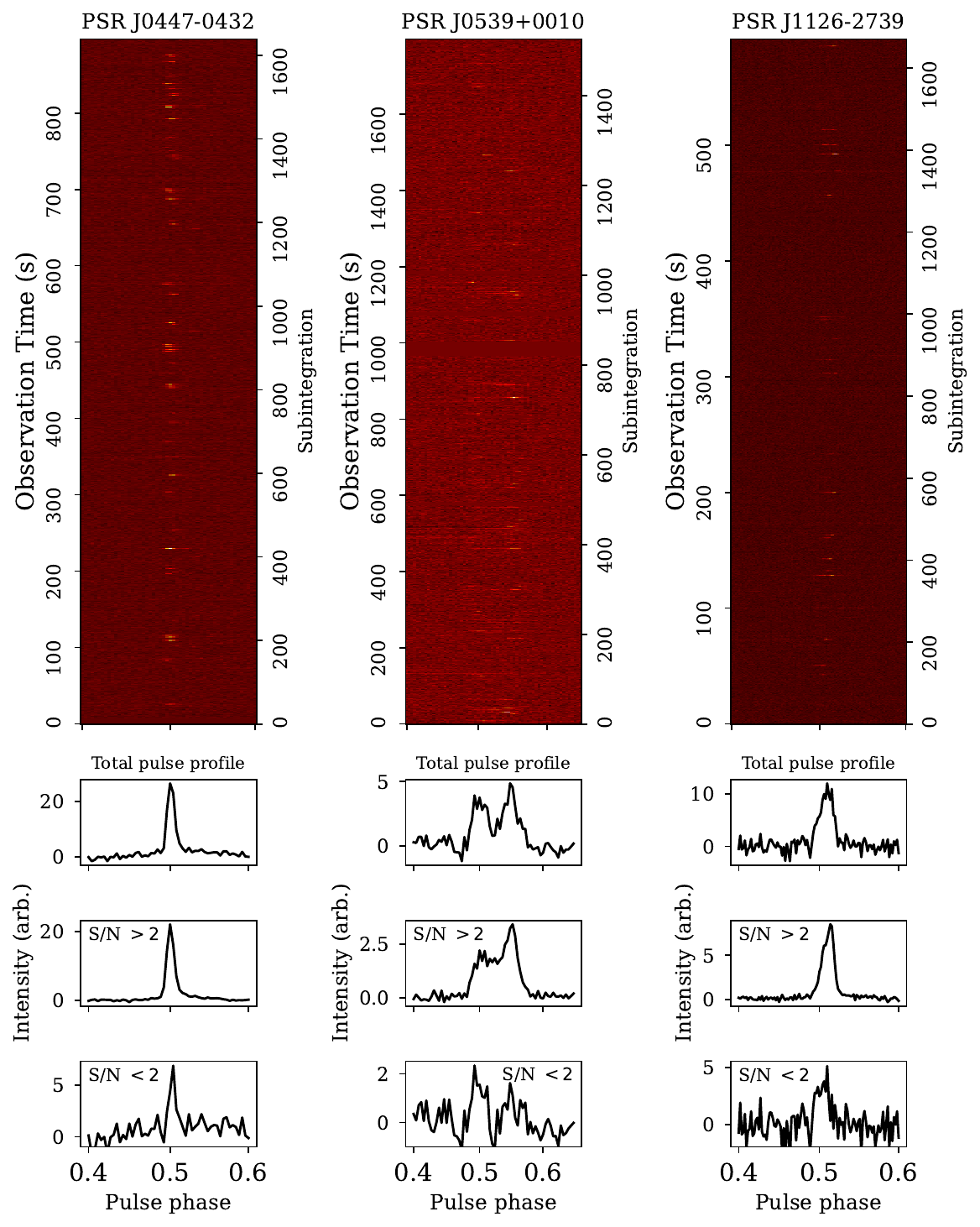}
    \caption{Characteristic observations of each pulsar. The top row displays a time-phase plot from one observation of each pulsar. Below, the total integrated profile, profile composed of only subintegrations with $\mathrm{S/N} > 2$, and profile composed of only subintegrations with $\mathrm{S/N} < 2$. Each plot covers only 10 to 20~$\%$ of a full rotation.}
    \label{fig:weak_emission}
\end{figure}

Characterization of the emission properties of many pulsars by the FAST GPPS survey grouped them into three loose groups: the majority of RRATs are ``normal'' pulsars with occasional bright pulses, some are extreme nullers, and some show very few bright pulses with no weak emission \citep{fast-gpps-rrats-23}. To determine the emission patterns of these pulsars, we used the de-dispersed data cubes to create time-phase plots and total integrated profiles for each observation. We then extracted all subintegrations with signal-to-noise ratios $\mathrm{S/N} > 2$, and created profiles exclusively from those pulses, as well as a ``noise'' profile consisting only of subintegrations with $\mathrm{S/N} < 2$, which are shown in Figure~\ref{fig:weak_emission}. While the majority of the emission comes from the sporadic, bright pulses, each pulsar shows evidence of sub-threshold emission, particularly J0447.

\section{Discussion}
\label{sec:disc}
\subsection{Timing}
\begin{figure*}
    \centering
    \includegraphics[width=\linewidth]{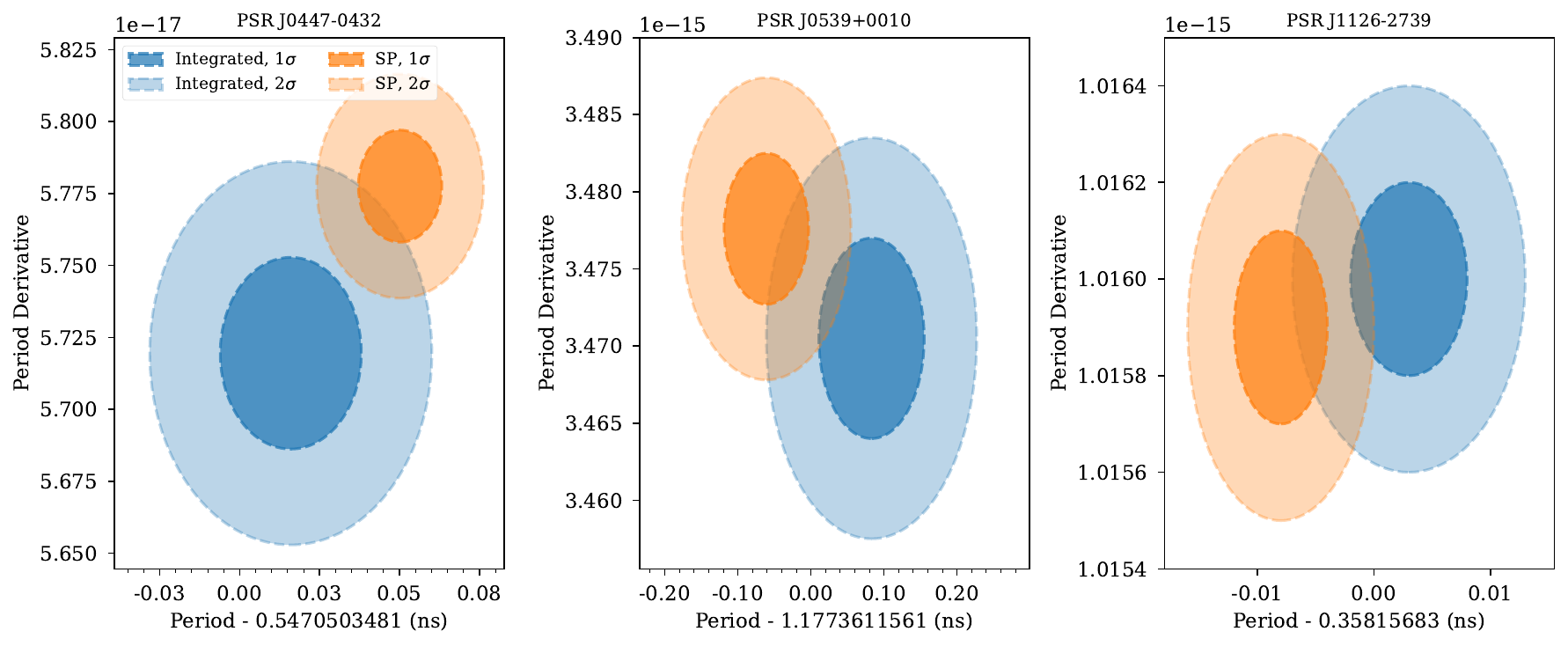}
    \caption{The period and period derivative found through pulsar timing for each pulsar, through integrated TOAs (blue) and single-pulse TOAs (orange). The 1-- and 2--$\sigma$ error bars on each parameter are denoted by the darker and lighter ellipses, respectively.}
    \label{fig:ppdot-error-ellipses}
\end{figure*}

\begin{figure*}
    \centering
    \includegraphics[width=\linewidth]{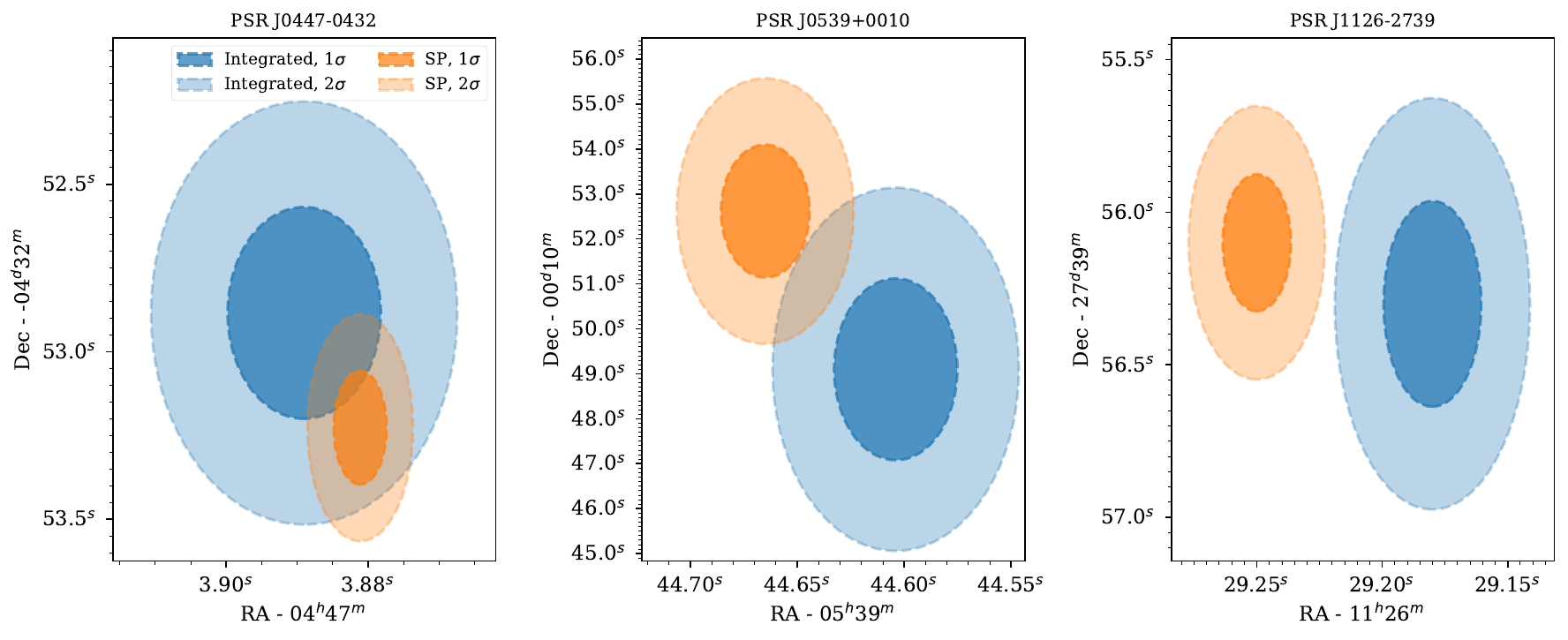}
    \caption{The right ascension and declination coordinates found through pulsar timing for each pulsar, through integrated TOAs (blue) and single-pulse TOAs (orange). The 1-- and 2--$\sigma$ error bars on each parameter are denoted by the darker and lighter ellipses, respectively.}
    \label{fig:sky-pos-error-ellipses}
\end{figure*}

We successfully obtained phase-connected timing solutions for these three pulsars using the standard method of integrating the emission by folding the data, as well as by measuring TOAs for each detected single pulse. The independently obtained sets of TOAs provide an opportunity to cross-reference the derived timing parameters between solutions, and evaluate the accuracy and precision of the two timing methods.

For all three pulsars, the single-pulse timing solutions have larger RMS post-fit residuals and larger reduced chi-squared values\footnote{Using the MODE 1 setting in \textsc{tempo2}, the uncertainties on the TOAs were multiplied by a factor so that the reduced chi-squared values of the final solutions would be one.}, suggesting that these timing models fit the observed TOAs less accurately, and that there would be more uncertainty in the derived parameters making up the timing model. However, the timing-derived parameters have error bars of similar sizes and are generally in good agreement between the two timing methods for all three pulsars. To visualize this, we have plotted the period and period derivative (Figure~\ref{fig:ppdot-error-ellipses}) and the sky positions (Figure~\ref{fig:sky-pos-error-ellipses}) returned by the best-fit timing models for each pulsar using both methods, as well as the error bars on each. All of the parameters are in agreement between the two solutions within two standard deviations. This indicates that the timing solutions are of comparable accuracy. The error bars on the parameters from single-pulse timing are consistently less than or equal to the error bars on the same parameters from the integrated timing solution; the greater number of TOAs and detection of single pulses on epochs without folding detections likely allowed for more accurate constraints on these parameters.

The two methods studied here yield spin parameters and sky positions with similar-sized error bars, yet the RMS and reduced chi squared ($\chi^2_r$) of the single-pulse solutions are significantly higher, due to the large number of TOAs scattered across a larger pulse phase envelope. To perform the fits, \textsc{tempo2} first measures the residual of each pulse, the difference between the predicted and observed arrival times based on the provided pre-fit timing model. \textsc{tempo2} uses a weighted least-squares algorithm to minimize the chi-squared value:

\begin{equation}
    \chi^2 = \sum_{i=1}^N  \left( \frac{R_i}{\sigma_i} \right)^2
\end{equation}

Here, $R_i$ is the residual of the $i^{th}$ TOA, $\sigma_i$ is the uncertainty on the $i^{th}$ TOA, and the sum is performed over all of the TOAs used in the timing model \citep{tempo2}. 

Under this formulation, a clear disparity arises between the two methods. Using folding, many single pulses which may arrive within a larger pulse envelope are averaged together into one pulse profile. On the other hand, each detected pulse may peak anywhere within a window that encompasses a few percent of the total rotational period of the source. These variations are intrinsic to the pulsar emission process, however, when these single-pulse TOAs are used to calculate a timing model, the large number of pulses with large residuals will yield a calculation of $\chi^2$ and the root-mean-square residual value which are much higher than the corresponding values in the integrated timing solution.

Hybrid timing models, formed from a combination of integrated and single-pulse TOAs, may be used for pulsars with infrequent observational cadences, or small amounts of TOAs that make timing difficult. Epochs with single-pulse TOAs but no measurable integrated profile are common in the RRAT-like emission regime, but we also have observations with integrated TOAs but no single-pulse TOAs. These are typically short observations, with a small number of detected single pulses with S/N $\leq 7$, too dim to return a useful TOA measurement. While a hybrid method would be useful for extending timing baselines where necessary, the different uncertainty calculations for different types of TOAs would require more careful analysis for the derived timing parameters and their errors.

\subsection{Pulse Jitter}
On short timescales, the uncertainty in TOA measurements are dominated by radiometer noise, jitter noise, and scintillation noise \citep{lma+19}. Radiometer noise typically dominates low S/N observations, as the limited sensitivity impacts the effectiveness of template matching. The single pulses from all pulsars vary stochastically in both shape and arrival phase, contributing an excess noise component known as pulse jitter. As observational sensitivity increases, pulse jitter is expected to become the dominant source of noise, especially for millisecond pulsars (\citealt{lma+19,mzk+23,www+24}).
Previous estimates of the jitter-induced uncertainty have come either from cross-correlation of the TOA errors from different frequency bands \citep{mzk+23}, or by comparing the observed variance of the TOAs with the variance of idealized simulated datasets for each pulsar (\citealt{pbs+21,www+24}), both of which are outside the scope of this work.

For each observational epoch, we calculate the mean TOA error of the integrated TOAs, and compare this to the RMS of all the single-pulse TOAs from that epoch. We only examined observations with more than five single-pulse TOAs, and calculated the ratio of the single-pulse RMS over the mean integrated TOA error. We find that the ratio varies between five and 25 for almost all of our observational epochs; averaging over all of the observations yields mean ratios of 19, 13, and eight for PSRs J0447, J0539, and J1126 respectively.

\subsection{Single-Pulse Properties}
In this work, we fitted log-normal and power-law functions to the normalized pulse energy distribution and found that power law distributions are preferred for two of the pulsars. Despite this, we see no evidence of giant pulses with energies $>10$ times the average pulse energy, and all three pulsars show evidence of weak emission outside of the bright pulses. We conclude that the emission mechanisms at play here are fundamentally the same as for standard pulsars.

We also analyzed the wait time distributions of all three pulsars and found that Weibull distributions were preferred, implying moderate amounts of pulse clustering in time. While some pulses arrive consecutively, we do not exclusively observe ``active'' periods of several consecutive pulses followed by periods with no emission, and we are able to detect weak emission outside the bright pulses for each pulsar, leading us to conclude that these pulsars are not just extremely nulling pulsars.


\section{Conclusion}
\label{sec:con}
We report phase-connected timing solutions for three radio pulsars discovered at 350 MHz with the GBNCC and GBT 350-MHz drift scan surveys. Two of these pulsars were initially published as RRATs due to their discovery through single-pulse detection pipelines, but in subsequent observations, were sufficiently bright to form integrated pulse profiles and measure times of arrival. We formed two distinct timing models from different sets of TOAs: the integrated pulse profiles, and the arrival times of all detected single pulses. The best-fit sky coordinates, period, and period derivative are in agreement within $\sim2.5$ standard deviations. The error bars on the parameters are of similar size between the two timing models, although the reported RMS error of the single-pulse solutions are much higher for two of the pulsars owing to the large number of single pulses falling within a relatively wide phase envelope. 

The wait-time distributions of each pulsar are better fit to a model in which pulses tend to be clustered together in time, suggesting that the pulses do not occur randomly or independently. The energy distributions tend to prefer power law distributions, although no pulses meet the S/N $>$ 10 threshold as a loose definition of ``giant pulses'', and the presence of weak emission outside the bright pulses indicates that these are weaker pulsars with sparse bright pulses, as is the case for many RRATs as they are observed with the next generation of more sensitive radio telescopes.
This work shows the applicability of multiple TOA generation methods, and highlights the variety of emission characteristics of pulsars under the RRAT umbrella.

\begin{acknowledgments}
E.F.L. and M.A.M. are supported by NSF award AST-2009425. M.A.M. is also supported by NSF Physics Frontiers Center award PHYS-2020265. The Green Bank Observatory is a facility of the National Science Foundation operated under cooperative agreement by Associated Universities, Inc. The West Virginia University Research Corporation supported some of the GBT observations uses for this study.
\end{acknowledgments}

\vspace{5mm}
\facilities{GBT (GUPPI)}
\software{
\texttt{Astropy} \citep[\url{https://www.astropy.org/};][]{astropy2022},
\texttt{Matplotlib} \citep[\url{https://matplotlib.org/};][]{matplotlib},
\texttt{Numpy} \citep[\url{https://numpy.org/};][]{numpy},
\texttt{PRESTO} \citep[\url{https://www.cv.nrao.edu/~sransom/presto/};][]{presto}, 
\texttt{pyGDSM} \citep[\url{https://github.com/telegraphic/pygdsm};][]{2016ascl.soft03013P},
\texttt{Scipy} \citep[\url{https://scipy.org/};][]{scipy},
\texttt{seaborn} \citep[\url{https://github.com/mwaskom/seaborn};][]{seaborn},
\texttt{TEMPO2} \citep[\url{https://www.atnf.csiro.au/research/pulsar/tempo2/};][]{tempo2}
}

\bibliography{biblio}{}
\bibliographystyle{aasjournal}

\end{document}

%% file: observations.tex
\begin{deluxetable*}{cccccccccc}
\label{tab:obs}
\tablecaption{Observations and Single Pulse Search Parameters}
\tablehead{\colhead{$F_c$} & \colhead{$\Delta F$} & \colhead{$t_{samp}$} & \colhead{PSR} & \colhead{$N_{obs}$} & \colhead{Downsampling} & \colhead{ $\Delta {\rm DM}$} & \colhead{${\rm DM}_{low}$} & \colhead{${\rm DM}_{high}$} \\
\colhead{(MHz)} & \colhead{(MHz)} & \colhead{($\mu$s)} & & \colhead{} & \colhead{Factor} & \colhead{(pc cm$^{-3}$)} & \colhead{(pc cm$^{-3}$)} & \colhead{(pc cm$^{-3}$)}}

\startdata
\multirow{3}*{350} & \multirow{3}*{100} & \multirow{3}*{81.92} & J0447--0432 & 8 & 1 & 0.01 & 25 & 35 \\
& & & J0539+0010 & 21 & 2 & 0.02 & 38.5 & 58.5 \\
& & & J1126--2739 & 21 & 1 & 0.01 & 22 & 32 \\
\hline
\multirow{2}*{820} & \multirow{2}*{200} & \multirow{2}*{40.96} & J0447--0432 & 14 & 1 & 0.05 & 25 & 35 \\
& & & J1126--2739 & 2 & 1 & 0.05 & 22 & 32 \\
\enddata
\tablecomments{Parameters for the GBT/GUPPI observations of each pulsar at each frequency. From left to right, we list the central observing frequency in MHz, the total observing bandwidth in MHz, the native sampling time, the pulsars observed at that frequency, the number of observations taken of that pulsar at that frequency, the downsampling factor used on the timeseries, the spacing between different DM trials, and the minimum and maximum dispersion measures searched. No observations of PSR J0539+0010 were taken at 820 MHz.} 

\end{deluxetable*}

%% file: 0447table.tex
\begin{deluxetable*}{ccc}
\tablewidth{0pt} 
\label{tab:0447}
\tablecaption{J0447--0432}
\tablehead{
\colhead{} & \colhead{Integrated Timing} & \colhead{Single Pulse Timing}
} 
\startdata 
Right ascension (J2000) & $04^{\rm{h}}\, 47^{\rm{m}}\, 03\, \fs 89(1)$ & $04^{\rm{h}}\, 47^{\rm{m}}\, 03\, \fs 881(4)$ \\
Declination (J2000) & $-04\arcdeg\, 32\arcmin\, 52\, \farcs9(3)$ & $-04\arcdeg\, 32\arcmin\, 53\, \farcs2(2)$ \\
Spin frequency (Hz) & 1.82798530966(7) & 1.82798530954(9)  \\
Spin frequency derivative (Hz s$^{-1}$) & --1.91(1) $\times 10^{-16}$ & --1.930(6) $\times 10^{-16}$ \\
Dispersion measure (pc cm$^{-3}$) & 29.85(5) & 29.75(1) \\ 
\hline
Spin period, $P$ (s) & 0.54705034812(2) & 0.54705034815(1)  \\
Spin-down rate, $\dot{P}$ (s s$^{-1}$) & 5.72(3) $\times 10^{-17}$ & 5.78(2) $\times 10^{-17}$ \\
Surface magnetic field, B ($10^{12}$ G) & \multicolumn{2}{c}{0.18} \\
Spin-down luminosity, $\dot{E}$ ($10^{31}$ erg s$^{-1}$) & \multicolumn{2}{c}{1.4} \\
Characteristic age, $\tau_c$ (Myr) & \multicolumn{2}{c}{150} \\
DM distance (kpc) (NE2001) & \multicolumn{2}{c}{1.3} \\
DM distance (kpc) (YMW16) & \multicolumn{2}{c}{0.7} \\
\hline
Data span (yr) & \multicolumn{2}{c}{3.14} \\
Start epoch (MJD) & \multicolumn{2}{c}{56758} \\
End epoch (MJD) & \multicolumn{2}{c}{57907} \\
Timing epoch & \multicolumn{2}{c}{57815} \\
Number of observations & 16 & 21  \\
Number of TOAs & 23 & 512  \\
Post-fit residual RMS ($\mu$s) & 596 & 1549 \\
Reduced chi-squared & 2.4 & 20.2
\enddata

\tablecomments{Timing parameters for the integrated and single-pulse timing models for PSR J0447--0432. 1$\sigma$ error bars are denoted with parenthesis on the last significant digit. The uncertainties have been adjusted with the MODE 1 setting in \textsc{tempo2} so that the final solution has a reduced chi-squared of one. Both timing models use the DE438 solar system ephemeris and are referenced to the TT(TAI) time standard.}
\end{deluxetable*}

%% file: 0539table.tex
\begin{deluxetable*}{ccc}
\tablewidth{0pt} 
\label{tab:0539}
\tablecaption{J0539+0010}
\tablehead{
\colhead{} & \colhead{Integrated Timing} & \colhead{Single Pulse Timing}
} 
\startdata 
Right ascension (J2000) & $05^{\rm{h}}\, 39^{\rm{m}}\, 44\, \fs 60(3)$ & $05^{\rm{h}}\, 39^{\rm{m}}\, 44\, \fs 66(2)$ \\
Declination (J2000) & $+00\arcdeg\, 10\arcmin\, 49\arcsec(2)$ & $+00\arcdeg\, 10\arcmin\, 52\arcsec(2)$ \\
Spin frequency (Hz) & 0.84935705136(5) & 0.84935705146(4) \\
Spin frequency derivative (Hz s$^{-1}$) & --2.504(5) $\times 10^{-15}$ & --2.509(3) $\times 10^{-15}$ \\
Dispersion measure (pc cm$^{-3}$) & \multicolumn{2}{c}{48.46(2)} \\ 
\hline
Spin period, $P$ (s) & 1.17736115618(7) & 1.17736115604(6) \\
Spin-down rate, $\dot{P}$ (s s$^{-1}$) & 3.470(7) $\times 10^{-15}$ & 3.478(5) $\times 10^{-15}$ \\
Surface magnetic field, B ($10^{12}$ G) & \multicolumn{2}{c}{2.0} \\
Spin-down luminosity, $\dot{E}$ ($10^{31}$ erg s$^{-1}$) & \multicolumn{2}{c}{8.4}  \\
Characteristic age, $\tau_c$ (Myr) & \multicolumn{2}{c}{5.4} \\
DM distance (kpc) (NE2001) & \multicolumn{2}{c}{2.0} \\
DM distance (kpc) (YMW16) & \multicolumn{2}{c}{1.4} \\
\hline
Data span (yr) & \multicolumn{2}{c}{1.82} \\
Start epoch (MJD) & \multicolumn{2}{c}{58109} \\
End epoch (MJD) & \multicolumn{2}{c}{58772} \\
Timing epoch & \multicolumn{2}{c}{58384} \\
Number of observations & 17 & 19 \\
Number of TOAs & 29 & 315 \\
RMS post-fit residuals ($\mu$s) & 1950 & 716 \\
Reduced chi-squared & 0.64 & 61
\enddata
\tablecomments{Timing parameters for the integrated and single-pulse timing models for PSR J0539+0010. 1$\sigma$ error bars are denoted with parenthesis on the last significant digit. The uncertainties have been adjusted with the MODE 1 setting in \textsc{tempo2} so that the final solution has a reduced chi-squared of one. Both timing models use the DE438 solar system ephemeris and are referenced to the TT(TAI) time standard.}
\end{deluxetable*}

%% file: 1126table.tex
\begin{deluxetable*}{ccc}
\tablewidth{0pt} 
\label{tab:1126}
\tablecaption{J1126--2739}
\tablehead{
\colhead{} & \colhead{Integrated Timing} & \colhead{Single-Pulse Timing}
} 
\startdata 
Right ascension (J2000) & $11^{\rm{h}}\, 26^{\rm{m}}\, 29\, \fs 18(2)$ & $11^{\rm{h}}\, 26^{\rm{m}}\, 29\, \fs 25(1)$ \\
Declination (J2000) & $-27\arcdeg\, 39\arcmin\, 56 \farcs 3(3)$ & $-27\arcdeg\, 39\arcmin\, 56 \farcs 1(2)$ \\
Spin frequency (Hz) & 2.79207295863(4) & 2.79207295872(3) \\
Spin frequency derivative (Hz s$^{-1}$) & --7.921(2) $\times 10^{-15}$ & --7.920(1) $\times 10^{-15}$ \\
Dispersion measure (pc cm$^{-3}$) &  \multicolumn{2}{c}{26.86(1)} \\ 
\hline
Spin period, $P$ (s) & 0.358156830004(4) & 0.358156829992(4) \\
Spin-down rate, $\dot{P}$ (s s$^{-1}$) & 1.0160(2) $\times 10^{-15}$ & 1.0159(2) $\times 10^{-15}$ \\
Surface magnetic field, B ($10^{12}$ G) & \multicolumn{2}{c}{0.610} \\
Spin-down luminosity, $\dot{E}$ ($10^{31}$ erg s$^{-1}$) & \multicolumn{2}{c}{87.3}  \\
Characteristic age, $\tau_c$ (Myr) & \multicolumn{2}{c}{5.6} \\
DM distance (kpc) (NE2001) & \multicolumn{2}{c}{1.1} \\
DM distance (kpc) (YMW16) & \multicolumn{2}{c}{0.4} \\
\hline
Data span (yr) & \multicolumn{2}{c}{4.7} \\
Start epoch (MJD) & \multicolumn{2}{c}{57153} \\
End epoch (MJD) & \multicolumn{2}{c}{58874} \\
Timing epoch & \multicolumn{2}{c}{58013} \\
Number of observations & 19 & 21 \\
Number of TOAs & 49 & 237 \\
RMS post-fit residuals ($\mu$s) & 1100 & 1742 \\
Reduced chi-squared & 1.91 & 54.8
\enddata
\tablecomments{Timing parameters for the integrated and single-pulse timing models for PSR J1126--2739. 1$\sigma$ error bars are denoted with parenthesis on the last significant digit. The uncertainties have been adjusted with the MODE 1 setting in \textsc{tempo2} so that the final solution has a reduced chi-squared of one. Both timing models use the DE438 solar system ephemeris and are referenced to the TT(TAI) time standard.}
\end{deluxetable*}

%% file: waittime_dists.tex
\begin{deluxetable}{c|c|cc|c}
\label{tab:wts}
\tablecaption{Wait Time Distribution Fit Parameters}

\tablehead{\colhead{PSR} & \colhead{$r_{P}$ (hr$^{-1}$)} & \colhead{$r_{W}$ (hr$^{-1}$)} & \colhead{$k_{W}$} & \colhead{FP}}

\startdata
J0447--0432 & $220\pm20$ & $130\pm40$ & $0.5\pm0.1$ & $3 \times 10^{-7}$ \\ 
J0539+0010  & $97\pm6$ & $56\pm7$ & $0.49\pm0.05$ & $9 \times 10^{-15}$ \\ 
J1126--2739 & $93\pm10$ & $58\pm15$ & $0.65\pm0.09$ & $7 \times 10^{-4}$ \\ 
\enddata

\tablecomments{The wait time distribution fit results: $r_P$ is the Poissonian burst rate, $r_W$ the Weibull burst rate, and $k_W$ the Weibull clustering parameter. The final column denotes the probability that the improvement to the fit from the Weibull distribution is due to random chance. 1-$\sigma$ error bars are denoted.}
\end{deluxetable}

%% file: biblio.bib
@ARTICLE{atnf,
       author = {{Manchester}, R.~N. and {Hobbs}, G.~B. and {Teoh}, A. and {Hobbs}, M.},
        title = "{The Australia Telescope National Facility Pulsar Catalogue}",
      journal = {\aj},
         year = 2005,
        month = apr,
       volume = {129},
       number = {4},
        pages = {1993-2006},
          doi = {10.1086/428488},
archivePrefix = {arXiv},
       eprint = {astro-ph/0412641},
 primaryClass = {astro-ph},
       adsurl = {https://ui.adsabs.harvard.edu/abs/2005AJ....129.1993M}
}

@ARTICLE{tempo2,
       author = {{Hobbs}, G.~B. and {Edwards}, R.~T. and {Manchester}, R.~N.},
        title = "{TEMPO2, a new pulsar-timing package - I. An overview}",
      journal = {\mnras},
         year = 2006,
        month = jun,
       volume = {369},
       number = {2},
        pages = {655-672},
          doi = {10.1111/j.1365-2966.2006.10302.x},
archivePrefix = {arXiv},
       eprint = {astro-ph/0603381},
 primaryClass = {astro-ph},
       adsurl = {https://ui.adsabs.harvard.edu/abs/2006MNRAS.369..655H}
}

@INPROCEEDINGS{guppi,
       author = {{DuPlain}, Ron and {Ransom}, Scott and {Demorest}, Paul and {Brandt}, Patrick and {Ford}, John and {Shelton}, Amy L.},
        title = "{Launching GUPPI: the Green Bank Ultimate Pulsar Processing Instrument}",
    booktitle = {Advanced Software and Control for Astronomy II},
         year = 2008,
       editor = {{Bridger}, Alan and {Radziwill}, Nicole M.},
       series = {Society of Photo-Optical Instrumentation Engineers (SPIE) Conference Series},
       volume = {7019},
        month = aug,
          eid = {70191D},
        pages = {70191D},
          doi = {10.1117/12.790003},
       adsurl = {https://ui.adsabs.harvard.edu/abs/2008SPIE.7019E..1DD}
}

@software{presto,
       author = {{Ransom}, Scott},
        title = "{PRESTO: PulsaR Exploration and Search TOolkit}",
 howpublished = {Astrophysics Source Code Library, record ascl:1107.017},
         year = 2011,
        month = jul,
          eid = {ascl:1107.017},
archivePrefix = {ascl},
       eprint = {1107.017},
       adsurl = {https://ui.adsabs.harvard.edu/abs/2011ascl.soft07017R}
}

@BOOK{hbopa,
       author = {{Lorimer}, D.~R. and {Kramer}, M.},
        title = "{Handbook of Pulsar Astronomy}",
         year = 2012,
       adsurl = {https://ui.adsabs.harvard.edu/abs/2012hpa..book.....L}
}

@ARTICLE{astropy2022,
       author = {{Astropy Collaboration} and {Price-Whelan}, Adrian M. and {Lim}, Pey Lian and {Earl}, Nicholas and {Starkman}, Nathaniel and {Bradley}, Larry and {Shupe}, David L. and {Patil}, Aarya A. and {Corrales}, Lia and {Brasseur}, C.~E. and {N{\"o}the}, Maximilian and {Donath}, Axel and {Tollerud}, Erik and {Morris}, Brett M. and {Ginsburg}, Adam and {Vaher}, Eero and {Weaver}, Benjamin A. and {Tocknell}, James and {Jamieson}, William and {van Kerkwijk}, Marten H. and {Robitaille}, Thomas P. and {Merry}, Bruce and {Bachetti}, Matteo and {G{\"u}nther}, H. Moritz and {Aldcroft}, Thomas L. and {Alvarado-Montes}, Jaime A. and {Archibald}, Anne M. and {B{\'o}di}, Attila and {Bapat}, Shreyas and {Barentsen}, Geert and {Baz{\'a}n}, Juanjo and {Biswas}, Manish and {Boquien}, M{\'e}d{\'e}ric and {Burke}, D.~J. and {Cara}, Daria and {Cara}, Mihai and {Conroy}, Kyle E. and {Conseil}, Simon and {Craig}, Matthew W. and {Cross}, Robert M. and {Cruz}, Kelle L. and {D'Eugenio}, Francesco and {Dencheva}, Nadia and {Devillepoix}, Hadrien A.~R. and {Dietrich}, J{\"o}rg P. and {Eigenbrot}, Arthur Davis and {Erben}, Thomas and {Ferreira}, Leonardo and {Foreman-Mackey}, Daniel and {Fox}, Ryan and {Freij}, Nabil and {Garg}, Suyog and {Geda}, Robel and {Glattly}, Lauren and {Gondhalekar}, Yash and {Gordon}, Karl D. and {Grant}, David and {Greenfield}, Perry and {Groener}, Austen M. and {Guest}, Steve and {Gurovich}, Sebastian and {Handberg}, Rasmus and {Hart}, Akeem and {Hatfield-Dodds}, Zac and {Homeier}, Derek and {Hosseinzadeh}, Griffin and {Jenness}, Tim and {Jones}, Craig K. and {Joseph}, Prajwel and {Kalmbach}, J. Bryce and {Karamehmetoglu}, Emir and {Ka{\l}uszy{\'n}ski}, Miko{\l}aj and {Kelley}, Michael S.~P. and {Kern}, Nicholas and {Kerzendorf}, Wolfgang E. and {Koch}, Eric W. and {Kulumani}, Shankar and {Lee}, Antony and {Ly}, Chun and {Ma}, Zhiyuan and {MacBride}, Conor and {Maljaars}, Jakob M. and {Muna}, Demitri and {Murphy}, N.~A. and {Norman}, Henrik and {O'Steen}, Richard and {Oman}, Kyle A. and {Pacifici}, Camilla and {Pascual}, Sergio and {Pascual-Granado}, J. and {Patil}, Rohit R. and {Perren}, Gabriel I. and {Pickering}, Timothy E. and {Rastogi}, Tanuj and {Roulston}, Benjamin R. and {Ryan}, Daniel F. and {Rykoff}, Eli S. and {Sabater}, Jose and {Sakurikar}, Parikshit and {Salgado}, Jes{\'u}s and {Sanghi}, Aniket and {Saunders}, Nicholas and {Savchenko}, Volodymyr and {Schwardt}, Ludwig and {Seifert-Eckert}, Michael and {Shih}, Albert Y. and {Jain}, Anany Shrey and {Shukla}, Gyanendra and {Sick}, Jonathan and {Simpson}, Chris and {Singanamalla}, Sudheesh and {Singer}, Leo P. and {Singhal}, Jaladh and {Sinha}, Manodeep and {Sip{\H{o}}cz}, Brigitta M. and {Spitler}, Lee R. and {Stansby}, David and {Streicher}, Ole and {{\v{S}}umak}, Jani and {Swinbank}, John D. and {Taranu}, Dan S. and {Tewary}, Nikita and {Tremblay}, Grant R. and {de Val-Borro}, Miguel and {Van Kooten}, Samuel J. and {Vasovi{\'c}}, Zlatan and {Verma}, Shresth and {de Miranda Cardoso}, Jos{\'e} Vin{\'\i}cius and {Williams}, Peter K.~G. and {Wilson}, Tom J. and {Winkel}, Benjamin and {Wood-Vasey}, W.~M. and {Xue}, Rui and {Yoachim}, Peter and {Zhang}, Chen and {Zonca}, Andrea and {Astropy Project Contributors}},
        title = "{The Astropy Project: Sustaining and Growing a Community-oriented Open-source Project and the Latest Major Release (v5.0) of the Core Package}",
      journal = {\apj},
         year = 2022,
        month = aug,
       volume = {935},
       number = {2},
          eid = {167},
        pages = {167},
          doi = {10.3847/1538-4357/ac7c74},
archivePrefix = {arXiv},
       eprint = {2206.14220},
 primaryClass = {astro-ph.IM},
       adsurl = {https://ui.adsabs.harvard.edu/abs/2022ApJ...935..167A}
}

@Article{matplotlib,
  Author    = {Hunter, J. D.},
  Title     = {Matplotlib: A 2D graphics environment},
  Journal   = {Computing in Science \& Engineering},
  Volume    = {9},
  Number    = {3},
  Pages     = {90--95},
  publisher = {IEEE COMPUTER SOC},
  doi       = {10.1109/MCSE.2007.55},
  year      = 2007
}

@Article{numpy,
 title         = {Array programming with {NumPy}},
 author        = {Charles R. Harris and K. Jarrod Millman and St{\'{e}}fan J.
                 van der Walt and Ralf Gommers and Pauli Virtanen and David
                 Cournapeau and Eric Wieser and Julian Taylor and Sebastian
                 Berg and Nathaniel J. Smith and Robert Kern and Matti Picus
                 and Stephan Hoyer and Marten H. van Kerkwijk and Matthew
                 Brett and Allan Haldane and Jaime Fern{\'{a}}ndez del
                 R{\'{i}}o and Mark Wiebe and Pearu Peterson and Pierre
                 G{\'{e}}rard-Marchant and Kevin Sheppard and Tyler Reddy and
                 Warren Weckesser and Hameer Abbasi and Christoph Gohlke and
                 Travis E. Oliphant},
 year          = {2020},
 month         = sep,
 journal       = {Nature},
 volume        = {585},
 number        = {7825},
 pages         = {357--362},
 doi           = {10.1038/s41586-020-2649-2},
 publisher     = {Springer Science and Business Media {LLC}},
 url           = {https://doi.org/10.1038/s41586-020-2649-2}
}

@MISC{2016ascl.soft03013P,
       author = {{Price}, Danny C.},
        title = "{PyGDSM: Python interface to Global Diffuse Sky Models}",
 howpublished = {Astrophysics Source Code Library, record ascl:1603.013},
         year = 2016,
        month = mar,
          eid = {ascl:1603.013},
       adsurl = {https://ui.adsabs.harvard.edu/abs/2016ascl.soft03013P}
}

@ARTICLE{scipy,
  author  = {Virtanen, Pauli and Gommers, Ralf and Oliphant, Travis E. and
            Haberland, Matt and Reddy, Tyler and Cournapeau, David and
            Burovski, Evgeni and Peterson, Pearu and Weckesser, Warren and
            Bright, Jonathan and {van der Walt}, St{\'e}fan J. and
            Brett, Matthew and Wilson, Joshua and Millman, K. Jarrod and
            Mayorov, Nikolay and Nelson, Andrew R. J. and Jones, Eric and
            Kern, Robert and Larson, Eric and Carey, C J and
            Polat, {\.I}lhan and Feng, Yu and Moore, Eric W. and
            {VanderPlas}, Jake and Laxalde, Denis and Perktold, Josef and
            Cimrman, Robert and Henriksen, Ian and Quintero, E. A. and
            Harris, Charles R. and Archibald, Anne M. and
            Ribeiro, Ant{\^o}nio H. and Pedregosa, Fabian and
            {van Mulbregt}, Paul and {SciPy 1.0 Contributors}},
  title   = {{{SciPy} 1.0: Fundamental Algorithms for Scientific
            Computing in Python}},
  journal = {Nature Methods},
  year    = {2020},
  volume  = {17},
  pages   = {261--272},
  adsurl  = {https://rdcu.be/b08Wh},
  doi     = {10.1038/s41592-019-0686-2},
}

@ARTICLE{seaborn,
       author = {{Waskom}, Michael},
        title = "{seaborn: statistical data visualization}",
      journal = {The Journal of Open Source Software},
         year = 2021,
        month = apr,
       volume = {6},
       number = {60},
          eid = {3021},
        pages = {3021},
          doi = {10.21105/joss.03021},
       adsurl = {https://ui.adsabs.harvard.edu/abs/2021JOSS....6.3021W}
}

@ARTICLE{g86,
       author = {{Gehrels}, N.},
        title = "{Confidence Limits for Small Numbers of Events in Astrophysical Data}",
      journal = {\apj},
         year = 1986,
        month = apr,
       volume = {303},
        pages = {336},
          doi = {10.1086/164079},
       adsurl = {https://ui.adsabs.harvard.edu/abs/1986ApJ...303..336G}
}

@ARTICLE{lkr+02,
       author = {{van Leeuwen}, A.~G.~J. and {Kouwenhoven}, M.~L.~A. and {Ramachandran}, R. and {Rankin}, J.~M. and {Stappers}, B.~W.},
        title = "{Null-induced mode changes in PSR B0809+74}",
      journal = {\aap},
         year = 2002,
        month = may,
       volume = {387},
        pages = {169-178},
          doi = {10.1051/0004-6361:20020254},
archivePrefix = {arXiv},
       eprint = {astro-ph/0202477},
 primaryClass = {astro-ph},
       adsurl = {https://ui.adsabs.harvard.edu/abs/2002A&A...387..169V}
}

@ARTICLE{cm03,
       author = {{Cordes}, J.~M. and {McLaughlin}, M.~A.},
        title = "{Searches for Fast Radio Transients}",
      journal = {\apj},
         year = 2003,
        month = oct,
       volume = {596},
       number = {2},
        pages = {1142-1154},
          doi = {10.1086/378231},
archivePrefix = {arXiv},
       eprint = {astro-ph/0304364},
 primaryClass = {astro-ph},
       adsurl = {https://ui.adsabs.harvard.edu/abs/2003ApJ...596.1142C}
}

@ARTICLE{mll+06,
       author = {{McLaughlin}, M.~A. and {Lyne}, A.~G. and {Lorimer}, D.~R. and {Kramer}, M. and {Faulkner}, A.~J. and {Manchester}, R.~N. and {Cordes}, J.~M. and {Camilo}, F. and {Possenti}, A. and {Stairs}, I.~H. and {Hobbs}, G. and {D'Amico}, N. and {Burgay}, M. and {O'Brien}, J.~T.},
        title = "{Transient radio bursts from rotating neutron stars}",
      journal = {\nat},
         year = 2006,
        month = feb,
       volume = {439},
       number = {7078},
        pages = {817-820},
          doi = {10.1038/nature04440},
archivePrefix = {arXiv},
       eprint = {astro-ph/0511587},
 primaryClass = {astro-ph},
       adsurl = {https://ui.adsabs.harvard.edu/abs/2006Natur.439..817M}
}

@ARTICLE{wmj07,
       author = {{Wang}, N. and {Manchester}, R.~N. and {Johnston}, S.},
        title = "{Pulsar nulling and mode changing}",
      journal = {\mnras},
         year = 2007,
        month = may,
       volume = {377},
       number = {3},
        pages = {1383-1392},
          doi = {10.1111/j.1365-2966.2007.11703.x},
archivePrefix = {arXiv},
       eprint = {astro-ph/0703241},
 primaryClass = {astro-ph},
       adsurl = {https://ui.adsabs.harvard.edu/abs/2007MNRAS.377.1383W}
}

@ARTICLE{kk08,
       author = {{Keane}, E.~F. and {Kramer}, M.},
        title = "{On the birthrates of Galactic neutron stars}",
      journal = {\mnras},
         year = 2008,
        month = dec,
       volume = {391},
       number = {4},
        pages = {2009-2016},
          doi = {10.1111/j.1365-2966.2008.14045.x},
archivePrefix = {arXiv},
       eprint = {0810.1512},
 primaryClass = {astro-ph},
       adsurl = {https://ui.adsabs.harvard.edu/abs/2008MNRAS.391.2009K}
}

@ARTICLE{mlk+09,
       author = {{McLaughlin}, M.~A. and {Lyne}, A.~G. and {Keane}, E.~F. and {Kramer}, M. and {Miller}, J.~J. and {Lorimer}, D.~R. and {Manchester}, R.~N. and {Camilo}, F. and {Stairs}, I.~H.},
        title = "{Timing observations of rotating radio transients}",
      journal = {\mnras},
         year = 2009,
        month = dec,
       volume = {400},
       number = {3},
        pages = {1431-1438},
          doi = {10.1111/j.1365-2966.2009.15584.x},
archivePrefix = {arXiv},
       eprint = {0908.3813},
 primaryClass = {astro-ph.SR},
       adsurl = {https://ui.adsabs.harvard.edu/abs/2009MNRAS.400.1431M}
}

@ARTICLE{lmk+09,
       author = {{Lyne}, A.~G. and {McLaughlin}, M.~A. and {Keane}, E.~F. and {Kramer}, M. and {Espinoza}, C.~M. and {Stappers}, B.~W. and {Palliyaguru}, N.~T. and {Miller}, J.},
        title = "{Unusual glitch activity in the RRAT J1819-1458: an exhausted magnetar?}",
      journal = {\mnras},
         year = 2009,
        month = dec,
       volume = {400},
       number = {3},
        pages = {1439-1444},
          doi = {10.1111/j.1365-2966.2009.15668.x},
archivePrefix = {arXiv},
       eprint = {0909.1165},
 primaryClass = {astro-ph.SR},
       adsurl = {https://ui.adsabs.harvard.edu/abs/2009MNRAS.400.1439L}
}

@ARTICLE{bb10,
       author = {{Burke-Spolaor}, S. and {Bailes}, M.},
        title = "{The millisecond radio sky: transients from a blind single-pulse search}",
      journal = {\mnras},
         year = 2010,
        month = feb,
       volume = {402},
       number = {2},
        pages = {855-866},
          doi = {10.1111/j.1365-2966.2009.15965.x},
archivePrefix = {arXiv},
       eprint = {0911.1790},
 primaryClass = {astro-ph.SR},
       adsurl = {https://ui.adsabs.harvard.edu/abs/2010MNRAS.402..855B}
}

@ARTICLE{kss+10,
       author = {{Karuppusamy}, R. and {Stappers}, B.~W. and {van Straten}, W.},
        title = "{Giant pulses from the Crab pulsar. A wide-band study}",
      journal = {\aap},
         year = 2010,
        month = jun,
       volume = {515},
          eid = {A36},
        pages = {A36},
          doi = {10.1051/0004-6361/200913729},
archivePrefix = {arXiv},
       eprint = {1004.2803},
 primaryClass = {astro-ph.GA},
       adsurl = {https://ui.adsabs.harvard.edu/abs/2010A&A...515A..36K}
}

@ARTICLE{km11,
       author = {{Keane}, E.~F. and {McLaughlin}, M.~A.},
        title = "{Rotating radio transients}",
      journal = {Bulletin of the Astronomical Society of India},
         year = 2011,
        month = sep,
       volume = {39},
       number = {3},
        pages = {333-352},
          doi = {10.48550/arXiv.1109.6896},
archivePrefix = {arXiv},
       eprint = {1109.6896},
 primaryClass = {astro-ph.SR},
       adsurl = {https://ui.adsabs.harvard.edu/abs/2011BASI...39..333K}
}

@ARTICLE{mml+12,
       author = {{Mickaliger}, M.~B. and {McLaughlin}, M.~A. and {Lorimer}, D.~R. and {Langston}, G.~I. and {Bilous}, A.~V. and {Kondratiev}, V.~I. and {Lyutikov}, M. and {Ransom}, S.~M. and {Palliyaguru}, N.},
        title = "{A Giant Sample of Giant Pulses from the Crab Pulsar}",
      journal = {\apj},
         year = 2012,
        month = nov,
       volume = {760},
       number = {1},
          eid = {64},
        pages = {64},
          doi = {10.1088/0004-637X/760/1/64},
archivePrefix = {arXiv},
       eprint = {1210.0452},
 primaryClass = {astro-ph.HE},
       adsurl = {https://ui.adsabs.harvard.edu/abs/2012ApJ...760...64M}
}

@ARTICLE{sjb+12,
       author = {{Burke-Spolaor}, S. and {Johnston}, S. and {Bailes}, M. and {Bates}, S.~D. and {Bhat}, N.~D.~R. and {Burgay}, M. and {Champion}, D.~J. and {D'Amico}, N. and {Keith}, M.~J. and {Kramer}, M. and {Levin}, L. and {Milia}, S. and {Possenti}, A. and {Stappers}, B. and {van Straten}, W.},
        title = "{The High Time Resolution Universe Pulsar Survey - V. Single-pulse energetics and modulation properties of 315 pulsars}",
      journal = {\mnras},
         year = 2012,
        month = jun,
       volume = {423},
       number = {2},
        pages = {1351-1367},
          doi = {10.1111/j.1365-2966.2012.20998.x},
archivePrefix = {arXiv},
       eprint = {1203.6068},
 primaryClass = {astro-ph.SR},
       adsurl = {https://ui.adsabs.harvard.edu/abs/2012MNRAS.423.1351B}
}

@ARTICLE{lbr+13,
       author = {{Lynch}, Ryan S. and {Boyles}, Jason and {Ransom}, Scott M. and {Stairs}, Ingrid H. and {Lorimer}, Duncan R. and {McLaughlin}, Maura A. and {Hessels}, Jason W.~T. and {Kaspi}, Victoria M. and {Kondratiev}, Vladislav I. and {Archibald}, Anne M. and {Berndsen}, Aaron and {Cardoso}, Rogerio F. and {Cherry}, Angus and {Epstein}, Courtney R. and {Karako-Argaman}, Chen and {McPhee}, Christie A. and {Pennucci}, Tim and {Roberts}, Mallory S.~E. and {Stovall}, Kevin and {van Leeuwen}, Joeri},
        title = "{The Green Bank Telescope 350 MHz Drift-scan Survey II: Data Analysis and the Timing of 10 New Pulsars, Including a Relativistic Binary}",
      journal = {\apj},
         year = 2013,
        month = feb,
       volume = {763},
       number = {2},
          eid = {81},
        pages = {81},
          doi = {10.1088/0004-637X/763/2/81},
archivePrefix = {arXiv},
       eprint = {1209.4296},
 primaryClass = {astro-ph.HE},
       adsurl = {https://ui.adsabs.harvard.edu/abs/2013ApJ...763...81L}
}

@ARTICLE{blr+13,
       author = {{Boyles}, J. and {Lynch}, R.~S. and {Ransom}, S.~M. and {Stairs}, I.~H. and {Lorimer}, D.~R. and {McLaughlin}, M.~A. and {Hessels}, J.~W.~T. and {Kaspi}, V.~M. and {Kondratiev}, V.~I. and {Archibald}, A. and {Berndsen}, A. and {Cardoso}, R.~F. and {Cherry}, A. and {Epstein}, C.~R. and {Karako-Argaman}, C. and {McPhee}, C.~A. and {Pennucci}, T. and {Roberts}, M.~S.~E. and {Stovall}, K. and {van Leeuwen}, J.},
        title = "{The Green Bank Telescope 350 MHz Drift-scan survey. I. Survey Observations and the Discovery of 13 Pulsars}",
      journal = {\apj},
         year = 2013,
        month = feb,
       volume = {763},
       number = {2},
          eid = {80},
        pages = {80},
          doi = {10.1088/0004-637X/763/2/80},
archivePrefix = {arXiv},
       eprint = {1209.4293},
 primaryClass = {astro-ph.HE},
       adsurl = {https://ui.adsabs.harvard.edu/abs/2013ApJ...763...80B}
}

@ARTICLE{slr+14,
       author = {{Stovall}, K. and {Lynch}, R.~S. and {Ransom}, S.~M. and {Archibald}, A.~M. and {Banaszak}, S. and {Biwer}, C.~M. and {Boyles}, J. and {Dartez}, L.~P. and {Day}, D. and {Ford}, A.~J. and {Flanigan}, J. and {Garcia}, A. and {Hessels}, J.~W.~T. and {Hinojosa}, J. and {Jenet}, F.~A. and {Kaplan}, D.~L. and {Karako-Argaman}, C. and {Kaspi}, V.~M. and {Kondratiev}, V.~I. and {Leake}, S. and {Lorimer}, D.~R. and {Lunsford}, G. and {Martinez}, J.~G. and {Mata}, A. and {McLaughlin}, M.~A. and {Roberts}, M.~S.~E. and {Rohr}, M.~D. and {Siemens}, X. and {Stairs}, I.~H. and {van Leeuwen}, J. and {Walker}, A.~N. and {Wells}, B.~L.},
        title = "{The Green Bank Northern Celestial Cap Pulsar Survey. I. Survey Description, Data Analysis, and Initial Results}",
      journal = {\apj},
         year = 2014,
        month = aug,
       volume = {791},
       number = {1},
          eid = {67},
        pages = {67},
          doi = {10.1088/0004-637X/791/1/67},
archivePrefix = {arXiv},
       eprint = {1406.5214},
 primaryClass = {astro-ph.HE},
       adsurl = {https://ui.adsabs.harvard.edu/abs/2014ApJ...791...67S}
}

@ARTICLE{KA15,
       author = {{Karako-Argaman}, C. and {Kaspi}, V.~M. and {Lynch}, R.~S. and {Hessels}, J.~W.~T. and {Kondratiev}, V.~I. and {McLaughlin}, M.~A. and {Ransom}, S.~M. and {Archibald}, A.~M. and {Boyles}, J. and {Jenet}, F.~A. and {Kaplan}, D.~L. and {Levin}, L. and {Lorimer}, D.~R. and {Madsen}, E.~C. and {Roberts}, M.~S.~E. and {Siemens}, X. and {Stairs}, I.~H. and {Stovall}, K. and {Swiggum}, J.~K. and {van Leeuwen}, J.},
        title = "{Discovery and Follow-up of Rotating Radio Transients with the Green Bank and LOFAR Telescopes}",
      journal = {\apj},
         year = 2015,
        month = aug,
       volume = {809},
       number = {1},
          eid = {67},
        pages = {67},
          doi = {10.1088/0004-637X/809/1/67},
archivePrefix = {arXiv},
       eprint = {1503.05170},
 primaryClass = {astro-ph.HE},
       adsurl = {https://ui.adsabs.harvard.edu/abs/2015ApJ...809...67K}
}

@ARTICLE{cbc+17,
       author = {{Cameron}, A.~D. and {Barr}, E.~D. and {Champion}, D.~J. and {Kramer}, M. and {Zhu}, W.~W.},
        title = "{An investigation of pulsar searching techniques with the fast folding algorithm}",
      journal = {\mnras},
         year = 2017,
        month = jun,
       volume = {468},
       number = {2},
        pages = {1994-2010},
          doi = {10.1093/mnras/stx589},
archivePrefix = {arXiv},
       eprint = {1703.05581},
 primaryClass = {astro-ph.IM},
       adsurl = {https://ui.adsabs.harvard.edu/abs/2017MNRAS.468.1994C}
}

@ARTICLE{cbm+17,
       author = {{Cui}, B.-Y. and {Boyles}, J. and {McLaughlin}, M.~A. and {Palliyaguru}, N.},
        title = "{Timing Solution and Single-pulse Properties for Eight Rotating Radio Transients}",
      journal = {\apj},
         year = 2017,
        month = may,
       volume = {840},
       number = {1},
          eid = {5},
        pages = {5},
          doi = {10.3847/1538-4357/aa6aa9},
archivePrefix = {arXiv},
       eprint = {1706.08412},
 primaryClass = {astro-ph.HE},
       adsurl = {https://ui.adsabs.harvard.edu/abs/2017ApJ...840....5C}
}

@ARTICLE{pkr+18,
       author = {{Parent}, E. and {Kaspi}, V.~M. and {Ransom}, S.~M. and {Krasteva}, M. and {Patel}, C. and {Scholz}, P. and {Brazier}, A. and {McLaughlin}, M.~A. and {Boyce}, M. and {Zhu}, W.~W. and {Pleunis}, Z. and {Allen}, B. and {Bogdanov}, S. and {Caballero}, K. and {Camilo}, F. and {Camuccio}, R. and {Chatterjee}, S. and {Cordes}, J.~M. and {Crawford}, F. and {Deneva}, J.~S. and {Ferdman}, R. and {Freire}, P.~C.~C. and {Hessels}, J.~W.~T. and {Jenet}, F.~A. and {Knispel}, B. and {Lazarus}, P. and {van Leeuwen}, J. and {Lyne}, A.~G. and {Lynch}, R. and {Seymour}, A. and {Siemens}, X. and {Stairs}, I.~H. and {Stovall}, K. and {Swiggum}, J.},
        title = "{The Implementation of a Fast-folding Pipeline for Long-period Pulsar Searching in the PALFA Survey}",
      journal = {\apj},
         year = 2018,
        month = jul,
       volume = {861},
       number = {1},
          eid = {44},
        pages = {44},
          doi = {10.3847/1538-4357/aac5f0},
archivePrefix = {arXiv},
       eprint = {1805.08247},
 primaryClass = {astro-ph.HE},
       adsurl = {https://ui.adsabs.harvard.edu/abs/2018ApJ...861...44P}
}

@ARTICLE{bsa+18,
       author = {{Shapiro-Albert}, B.~J. and {McLaughlin}, M.~A. and {Keane}, E.~F.},
        title = "{Radio Properties of Rotating Radio Transients: Single-pulse Spectral and Wait-time Analyses}",
      journal = {\apj},
         year = 2018,
        month = oct,
       volume = {866},
       number = {2},
          eid = {152},
        pages = {152},
          doi = {10.3847/1538-4357/aae2b2},
archivePrefix = {arXiv},
       eprint = {1809.06729},
 primaryClass = {astro-ph.HE},
       adsurl = {https://ui.adsabs.harvard.edu/abs/2018ApJ...866..152S}
}

@ARTICLE{jvk+18,
       author = {{Jankowski}, F. and {van Straten}, W. and {Keane}, E.~F. and {Bailes}, M. and {Barr}, E.~D. and {Johnston}, S. and {Kerr}, M.},
        title = "{Spectral properties of 441 radio pulsars}",
      journal = {\mnras},
         year = 2018,
        month = feb,
       volume = {473},
       number = {4},
        pages = {4436-4458},
          doi = {10.1093/mnras/stx2476},
archivePrefix = {arXiv},
       eprint = {1709.08864},
 primaryClass = {astro-ph.HE},
       adsurl = {https://ui.adsabs.harvard.edu/abs/2018MNRAS.473.4436J}
}

@ARTICLE{lsk+18,
       author = {{Lynch}, Ryan S. and {Swiggum}, Joseph K. and {Kondratiev}, Vlad I. and {Kaplan}, David L. and {Stovall}, Kevin and {Fonseca}, Emmanuel and {Roberts}, Mallory S.~E. and {Levin}, Lina and {DeCesar}, Megan E. and {Cui}, Bingyi and {Cenko}, S. Bradley and {Gatkine}, Pradip and {Archibald}, Anne M. and {Banaszak}, Shawn and {Biwer}, Christopher M. and {Boyles}, Jason and {Chawla}, Pragya and {Dartez}, Louis P. and {Day}, David and {Ford}, Anthony J. and {Flanigan}, Joseph and {Hessels}, Jason W.~T. and {Hinojosa}, Jesus and {Jenet}, Fredrick A. and {Karako-Argaman}, Chen and {Kaspi}, Victoria M. and {Leake}, Sean and {Lunsford}, Grady and {Martinez}, Jos{\'e} G. and {Mata}, Alberto and {McLaughlin}, Maura A. and {Noori}, Hind Al and {Ransom}, Scott M. and {Rohr}, Matthew D. and {Siemens}, Xavier and {Spiewak}, Ren{\'e}e and {Stairs}, Ingrid H. and {van Leeuwen}, Joeri and {Walker}, Arielle N. and {Wells}, Bradley L.},
        title = "{The Green Bank North Celestial Cap Pulsar Survey. III. 45 New Pulsar Timing Solutions}",
      journal = {\apj},
         year = 2018,
        month = jun,
       volume = {859},
       number = {2},
          eid = {93},
        pages = {93},
          doi = {10.3847/1538-4357/aabf8a},
archivePrefix = {arXiv},
       eprint = {1805.04951},
 primaryClass = {astro-ph.HE},
       adsurl = {https://ui.adsabs.harvard.edu/abs/2018ApJ...859...93L}
}

@ARTICLE{kmk+18,
       author = {{Kawash}, A.~M. and {McLaughlin}, M.~A. and {Kaplan}, D.~L. and {DeCesar}, M.~E. and {Levin}, L. and {Lorimer}, D.~R. and {Lynch}, R.~S. and {Stovall}, K. and {Swiggum}, J.~K. and {Fonseca}, E. and {Archibald}, A.~M. and {Banaszak}, S. and {Biwer}, C.~M. and {Boyles}, J. and {Cui}, B. and {Dartez}, L.~P. and {Day}, D. and {Ernst}, S. and {Ford}, A.~J. and {Flanigan}, J. and {Heatherly}, S.~A. and {Hessels}, J.~W.~T. and {Hinojosa}, J. and {Jenet}, F.~A. and {Karako-Argaman}, C. and {Kaspi}, V.~M. and {Kondratiev}, V.~I. and {Leake}, S. and {Lunsford}, G. and {Martinez}, J.~G. and {Mata}, A. and {Matheny}, T.~D. and {Mcewen}, A.~E. and {Mingyar}, M.~G. and {Orsini}, A.~L. and {Ransom}, S.~M. and {Roberts}, M.~S.~E. and {Rohr}, M.~D. and {Siemens}, X. and {Spiewak}, R. and {Stairs}, I.~H. and {van Leeuwen}, J. and {Walker}, A.~N. and {Wells}, B.~L.},
        title = "{The Green Bank Northern Celestial Cap Pulsar Survey. II. The Discovery and Timing of 10 Pulsars}",
      journal = {\apj},
         year = 2018,
        month = apr,
       volume = {857},
       number = {2},
          eid = {131},
        pages = {131},
          doi = {10.3847/1538-4357/aab61d},
archivePrefix = {arXiv},
       eprint = {1803.03587},
 primaryClass = {astro-ph.HE},
       adsurl = {https://ui.adsabs.harvard.edu/abs/2018ApJ...857..131K}
}

@ARTICLE{crafts,
       author = {{Li}, Di and {Wang}, Pei and {Qian}, Lei and {Krco}, Marko and {Jiang}, Peng and {Yue}, Youling and {Jin}, Chenjin and {Zhu}, Yan and {Pan}, Zhichen and {Nan}, Rendong and {Dunning}, Alex},
        title = "{FAST in Space: Considerations for a Multibeam, Multipurpose Survey Using China's 500-m Aperture Spherical Radio Telescope (FAST)}",
      journal = {IEEE Microwave Magazine},
         year = 2018,
        month = apr,
       volume = {19},
       number = {3},
        pages = {112-119},
          doi = {10.1109/MMM.2018.2802178},
archivePrefix = {arXiv},
       eprint = {1802.03709},
 primaryClass = {astro-ph.IM},
       adsurl = {https://ui.adsabs.harvard.edu/abs/2018IMMag..19..112L}
}

@ARTICLE{mmm+18,
       author = {{Mickaliger}, Mitchell B. and {McEwen}, A.~E. and {McLaughlin}, M.~A. and {Lorimer}, D.~R.},
        title = "{A study of single pulses in the Parkes Multibeam Pulsar Survey}",
      journal = {\mnras},
         year = 2018,
        month = oct,
       volume = {479},
       number = {4},
        pages = {5413-5422},
          doi = {10.1093/mnras/sty1785},
archivePrefix = {arXiv},
       eprint = {1807.00143},
 primaryClass = {astro-ph.HE},
       adsurl = {https://ui.adsabs.harvard.edu/abs/2018MNRAS.479.5413M}
}

@ARTICLE{oppermann+18,
       author = {{Oppermann}, Niels and {Yu}, Hao-Ran and {Pen}, Ue-Li},
        title = "{On the non-Poissonian repetition pattern of FRB121102}",
      journal = {\mnras},
         year = 2018,
        month = apr,
       volume = {475},
       number = {4},
        pages = {5109-5115},
          doi = {10.1093/mnras/sty004},
archivePrefix = {arXiv},
       eprint = {1705.04881},
 primaryClass = {astro-ph.HE},
       adsurl = {https://ui.adsabs.harvard.edu/abs/2018MNRAS.475.5109O}
}

@ARTICLE{acd+19,
       author = {{Aloisi}, R.~J. and {Cruz}, A. and {Daniels}, L. and {Meyers}, N. and {Roekle}, R. and {Schuett}, A. and {Swiggum}, J.~K. and {DeCesar}, M.~E. and {Kaplan}, D.~L. and {Lynch}, R.~S. and {Stovall}, K. and {Levin}, Lina and {Archibald}, A.~M. and {Banaszak}, S. and {Biwer}, C.~M. and {Boyles}, J. and {Chawla}, P. and {Dartez}, L.~P. and {Cui}, B. and {Day}, D.~F. and {Ford}, A.~J. and {Flanigan}, J. and {Fonseca}, E. and {Hessels}, J.~W.~T. and {Hinojosa}, J. and {Karako-Argaman}, C. and {Kaspi}, V.~M. and {Kondratiev}, V.~I. and {Leake}, S. and {Lunsford}, G. and {Martinez}, J.~G. and {Mata}, A. and {McLaughlin}, M.~A. and {Noori}, H. Al and {Ransom}, S.~M. and {Roberts}, M.~S.~E. and {Rohr}, M.~D. and {Siemens}, X. and {Spiewak}, R. and {Stairs}, I.~H. and {van Leeuwen}, J. and {Walker}, A.~N. and {Wells}, B.~L.},
        title = "{The Green Bank North Celestial Cap Pulsar Survey. IV. Four New Timing Solutions}",
      journal = {\apj},
         year = 2019,
        month = apr,
       volume = {875},
       number = {1},
          eid = {19},
        pages = {19},
          doi = {10.3847/1538-4357/ab0d21},
archivePrefix = {arXiv},
       eprint = {1903.03543},
 primaryClass = {astro-ph.HE},
       adsurl = {https://ui.adsabs.harvard.edu/abs/2019ApJ...875...19A}
}

@ARTICLE{lma+19,
       author = {{Lam}, M.~T. and {McLaughlin}, M.~A. and {Arzoumanian}, Z. and {Blumer}, H. and {Brook}, P.~R. and {Cromartie}, H.~T. and {Demorest}, P.~B. and {DeCesar}, M.~E. and {Dolch}, T. and {Ellis}, J.~A. and {Ferdman}, R.~D. and {Ferrara}, E.~C. and {Fonseca}, E. and {Garver-Daniels}, N. and {Gentile}, P.~A. and {Jones}, M.~L. and {Lorimer}, D.~R. and {Lynch}, R.~S. and {Ng}, C. and {Nice}, D.~J. and {Pennucci}, T.~T. and {Ransom}, S.~M. and {Spiewak}, R. and {Stairs}, I.~H. and {Stovall}, K. and {Swiggum}, J.~K. and {Vigeland}, S.~J. and {Zhu}, W.~W.},
        title = "{The NANOGrav 12.5 yr Data Set: The Frequency Dependence of Pulse Jitter in Precision Millisecond Pulsars}",
      journal = {\apj},
         year = 2019,
        month = feb,
       volume = {872},
       number = {2},
          eid = {193},
        pages = {193},
          doi = {10.3847/1538-4357/ab01cd},
archivePrefix = {arXiv},
       eprint = {1809.03058},
 primaryClass = {astro-ph.HE},
       adsurl = {https://ui.adsabs.harvard.edu/abs/2019ApJ...872..193L}
}

@ARTICLE{lmk+20,
       author = {{van Leeuwen}, Joeri and {Mikhailov}, Klim and {Keane}, Evan and {Coenen}, Thijs and {Connor}, Liam and {Kondratiev}, Vlad and {Michilli}, Daniele and {Sanidas}, Sotiris},
        title = "{LOFAR radio search for single and periodic pulses from M 31}",
      journal = {\aap},
         year = 2020,
        month = feb,
       volume = {634},
          eid = {A3},
        pages = {A3},
          doi = {10.1051/0004-6361/201937065},
archivePrefix = {arXiv},
       eprint = {1911.11228},
 primaryClass = {astro-ph.HE},
       adsurl = {https://ui.adsabs.harvard.edu/abs/2020A&A...634A...3V}
}

@ARTICLE{mss+20,
       author = {{McEwen}, A.~E. and {Spiewak}, R. and {Swiggum}, J.~K. and {Kaplan}, D.~L. and {Fiore}, W. and {Agazie}, G.~Y. and {Blumer}, H. and {Chawla}, P. and {DeCesar}, M. and {Kaspi}, V.~M. and {Kondratiev}, V.~I. and {LaRose}, M. and {Levin}, L. and {Lynch}, R.~S. and {McLaughlin}, M. and {Mingyar}, M. and {Noori}, H. Al and {Ransom}, S.~M. and {Roberts}, M.~S.~E. and {Schmiedekamp}, A. and {Schmiedekamp}, C. and {Siemens}, X. and {Stairs}, I. and {Stovall}, K. and {Surnis}, M. and {van Leeuwen}, J.},
        title = "{The Green Bank North Celestial Cap Pulsar Survey. V. Pulsar Census and Survey Sensitivity}",
      journal = {\apj},
         year = 2020,
        month = apr,
       volume = {892},
       number = {2},
          eid = {76},
        pages = {76},
          doi = {10.3847/1538-4357/ab75e2},
archivePrefix = {arXiv},
       eprint = {1909.11109},
 primaryClass = {astro-ph.HE},
       adsurl = {https://ui.adsabs.harvard.edu/abs/2020ApJ...892...76M}
}

@ARTICLE{syw+21,
       author = {{Sun}, Sheng-Nan and {Yan}, Wen-Ming and {Wang}, Na and {Yuen}, Rai},
        title = "{A single pulse study of PSR J1752+2359}",
      journal = {Research in Astronomy and Astrophysics},
         year = 2021,
        month = nov,
       volume = {21},
       number = {9},
          eid = {240},
        pages = {240},
          doi = {10.1088/1674-4527/21/9/240},
       adsurl = {https://ui.adsabs.harvard.edu/abs/2021RAA....21..240S}
}

@ARTICLE{pbs+21,
       author = {{Parthasarathy}, A. and {Bailes}, M. and {Shannon}, R.~M. and {van Straten}, W. and {Os{\l}owski}, S. and {Johnston}, S. and {Spiewak}, R. and {Reardon}, D.~J. and {Kramer}, M. and {Venkatraman Krishnan}, V. and {Pennucci}, T.~T. and {Abbate}, F. and {Buchner}, S. and {Camilo}, F. and {Champion}, D.~J. and {Geyer}, M. and {Hugo}, B. and {Jameson}, A. and {Karastergiou}, A. and {Keith}, M.~J. and {Serylak}, M.},
        title = "{Measurements of pulse jitter and single-pulse variability in millisecond pulsars using MeerKAT}",
      journal = {\mnras},
         year = 2021,
        month = mar,
       volume = {502},
       number = {1},
        pages = {407-422},
          doi = {10.1093/mnras/stab037},
archivePrefix = {arXiv},
       eprint = {2101.08531},
 primaryClass = {astro-ph.HE},
       adsurl = {https://ui.adsabs.harvard.edu/abs/2021MNRAS.502..407P}
}

@ARTICLE{abhishek22,
       author = {{Abhishek} and {Malusare}, Namrata and {Tanushree}, N. and {Hegde}, Gayathri and {Konar}, Sushan},
        title = "{Radio pulsar sub-populations (II): The mysterious RRATs}",
      journal = {Journal of Astrophysics and Astronomy},
         year = 2022,
        month = dec,
       volume = {43},
       number = {2},
          eid = {75},
        pages = {75},
          doi = {10.1007/s12036-022-09862-3},
archivePrefix = {arXiv},
       eprint = {2201.00295},
 primaryClass = {astro-ph.HE},
       adsurl = {https://ui.adsabs.harvard.edu/abs/2022JApA...43...75A}
}

@ARTICLE{bbc+22,
       author = {{Bezuidenhout}, M.~C. and {Barr}, E. and {Caleb}, M. and {Driessen}, L.~N. and {Jankowski}, F. and {Kramer}, M. and {Malenta}, M. and {Morello}, V. and {Rajwade}, K. and {Sanidas}, S. and {Stappers}, B.~W. and {Surnis}, M.},
        title = "{MeerTRAP: 12 Galactic fast transients detected in a real-time, commensal MeerKAT survey}",
      journal = {\mnras},
         year = 2022,
        month = may,
       volume = {512},
       number = {1},
        pages = {1483-1498},
          doi = {10.1093/mnras/stac579},
archivePrefix = {arXiv},
       eprint = {2203.00557},
 primaryClass = {astro-ph.HE},
       adsurl = {https://ui.adsabs.harvard.edu/abs/2022MNRAS.512.1483B}
}

@ARTICLE{mzk+23,
       author = {{Miao}, X.~L. and {Zhu}, W.~W. and {Kramer}, M. and {Freire}, P.~C.~C. and {Shao}, L. and {Yuan}, M. and {Meng}, L.~Q. and {Wu}, Z.~W. and {Miao}, C.~C. and {Guo}, Y.~J. and {Champion}, D.~J. and {Fonseca}, E. and {Yao}, J.~M. and {Xue}, M.~Y. and {Niu}, J.~R. and {Hu}, H. and {Zhang}, C.~M.},
        title = "{Variability, polarimetry, and timing properties of single pulses from PSR J2222-0137 using FAST}",
      journal = {\mnras},
         year = 2023,
        month = dec,
       volume = {526},
       number = {2},
        pages = {2156-2166},
          doi = {10.1093/mnras/stad2595},
archivePrefix = {arXiv},
       eprint = {2308.10683},
 primaryClass = {astro-ph.HE},
       adsurl = {https://ui.adsabs.harvard.edu/abs/2023MNRAS.526.2156M}
}

@ARTICLE{fast-gpps-rrats-23,
       author = {{Zhou}, D.~J. and {Han}, J.~L. and {Xu}, Jun and {Wang}, Chen and {Wang}, P.~F. and {Wang}, Tao and {Jing}, Wei-Cong and {Chen}, Xue and {Yan}, Yi and {Su}, Wei-Qi. and {Gan}, Heng-Qian and {Jiang}, Peng and {Sun}, Jing-Hai and {Wang}, Hong-Guang and {Wang}, Na and {Wang}, Shuang-Qiang and {Xu}, Ren-Xin and {You}, Xiao-Peng},
        title = "{The FAST Galactic Plane Pulsar Snapshot Survey. II. Discovery of 76 Galactic Rotating Radio Transients and the Enigma of RRATs}",
      journal = {Research in Astronomy and Astrophysics},
         year = 2023,
        month = oct,
       volume = {23},
       number = {10},
          eid = {104001},
        pages = {104001},
          doi = {10.1088/1674-4527/accc76},
archivePrefix = {arXiv},
       eprint = {2303.17279},
 primaryClass = {astro-ph.HE},
       adsurl = {https://ui.adsabs.harvard.edu/abs/2023RAA....23j4001Z}
}

@ARTICLE{dbh+24,
       author = {{Doskoch}, Graham M. and {Basuroski}, Andrea and {Halley}, Kriisa and {Sookram}, Avinash and {Rodriguez-Ramos}, Iliomar and {Nahata}, Valmik and {Rahman}, Zahi and {Zhang}, Maureen and {Uhlmann}, Ashish and {Lynch}, Abby and {Lewandowska}, Natalia and {Miranda}, Nohely and {Schmiedekamp}, Ann and {Schmiedekamp}, Carl and {McLaughlin}, Maura A. and {Reichart}, Daniel E. and {Haislip}, Joshua B. and {Kouprianov}, Vladimir V. and {White}, Steve and {Ghigo}, Frank and {Heatherly}, Sue Ann},
        title = "{A Statistical Analysis of Crab Pulsar Giant Pulse Rates}",
      journal = {\apj},
         year = 2024,
        month = oct,
       volume = {973},
       number = {2},
          eid = {87},
        pages = {87},
          doi = {10.3847/1538-4357/ad6304},
archivePrefix = {arXiv},
       eprint = {2407.15996},
 primaryClass = {astro-ph.HE},
       adsurl = {https://ui.adsabs.harvard.edu/abs/2024ApJ...973...87D}
}

@ARTICLE{www+24,
       author = {{Wang}, S.~Q. and {Wang}, N. and {Wang}, J.~B. and {Hobbs}, G. and {Xu}, H. and {Wang}, B.~J. and {Dai}, S. and {Dang}, S.~J. and {Li}, D. and {Feng}, Y. and {Zhang}, C.~M.},
        title = "{Pulse Jitter and Single-pulse Variability in Millisecond Pulsars}",
      journal = {\apj},
         year = 2024,
        month = mar,
       volume = {964},
       number = {1},
          eid = {6},
        pages = {6},
          doi = {10.3847/1538-4357/ad217b},
archivePrefix = {arXiv},
       eprint = {2401.12426},
 primaryClass = {astro-ph.HE},
       adsurl = {https://ui.adsabs.harvard.edu/abs/2024ApJ...964....6W}
}

@ARTICLE{gly+25,
       author = {{Gao}, Shi-Jie and {Li}, Xiang-Dong and {Yan}, Zhen and {Shao}, Yi-Xuan and {Zhou}, Ping},
        title = "{RRAT J2325{\textendash}0530: A Rotating Radio Transient with an Atypical Waiting-time Distribution}",
      journal = {\apj},
         year = 2025,
        month = oct,
       volume = {991},
       number = {2},
          eid = {201},
        pages = {201},
          doi = {10.3847/1538-4357/adff5b},
archivePrefix = {arXiv},
       eprint = {2508.17657},
 primaryClass = {astro-ph.HE},
       adsurl = {https://ui.adsabs.harvard.edu/abs/2025ApJ...991..201G}
}
